\documentclass[aps, prb, epsf, twocolumn, showpacs, floatfix]{revtex4-2}
\usepackage{CJK, appendix, amssymb, graphics,epsfig, epstopdf, color, verbatim, ulem, braket, tabularx, float, subfigure, syntonly}
\usepackage[dvipsnames,cmyk]{xcolor}
\usepackage{booktabs}
\usepackage[misc]{ifsym}
\usepackage[version=4]{mhchem}
\usepackage[colorlinks,linkcolor=blue,citecolor=blue,urlcolor=blue,anchorcolor=blue]{hyperref}
\usepackage{IEEEtrantools, booktabs, multirow,  colortbl, mathrsfs, longtable}

\begin{document}
\begin{CJK*}{GBK}{song}
\title{\mbox{Magnon, doublon and quarton excitations in 2D $S$=1/2 trimerized Heisenberg models} }
\author{Yue-Yue Chang$^{1}$}
\author{Jun-Qing Cheng$^{2}$}
\email{chengjq@gbu.edu.cn}
\author{Hui Shao$^{3,4}$}
\author{Dao-Xin Yao$^{1,5}$}
\email{yaodaox@mail.sysu.edu.cn}
\author{Han-Qing Wu$^{1}$}
\email{wuhanq3@mail.sysu.edu.cn}
\affiliation{\mbox{$^{1}$Guangdong Provincial Key Laboratory of Magnetoelectric Physics and Devices,}
\mbox{State Key Laboratory of Optoelectronic Materials and Technologies, Center for Neutron Science and Technology,}
\mbox{School of Physics, Sun Yat-sen University, Guangzhou, 510275, China}
\mbox{$^{2}$School of Physical Sciences, Great Bay University, Dongguan 523000, China,} 
\mbox{ and Great Bay Institute for Advanced Study, Dongguan 523000, China}
\mbox{$^{3}$Center for Advanced Quantum Studies, Department of Physics, Beijing Normal University, Beijing 100875, China}
\mbox{$^{4}$Key Laboratory of Multiscale Spin Physics, Ministry of Education, Beijing 100875, China}
\mbox{$^{5}$International Quantum Academy, Shenzhen 518048, China}}
\begin{abstract}
We investigate the magnetic excitations of the 2D $S$=1/2 trimerized Heisenberg models with intratrimer interaction $J_1$ and intertrimer interaction $J_2$ on four different lattices using a combination of stochastic series expansion quantum Monte Carlo (SSE QMC) and stochastic analytic continuation methods (SAC), complemented by cluster perturbation theory (CPT). These models exhibit quasi-particle-like excitations when $g=J_2/J_1$ is weak, characterized by low-energy magnons, intermediate-energy doublons, and high-energy quartons. The low-energy magnons are associated with the magnetic ground states. They can be described by the linear spin wave theory (LSWT) of the effective block spin model and the original spin model. Doublons and quartons emerge from the corresponding internal excitations of the trimers with distinct energy levels, which can be effectively analyzed using perturbative calculation when the ratio of exchange interactions $g$ is weak. In this weak $g$ regime, we observe a clear separation between the magnon and higher-energy spectra. As $g$ increases, doublon and quarton gradually merge into the magnon modes or some continua. Notably, in the Collinear II and trimerized Hexagon lattice, a broad continuum emerges above the single-magnon spectrum, originating from the quasi-1D physics due to the dilute connections between chains. In addition, we also compare our numerical results to the experimental RIXS spectrum and analyze the difference. Our numerical analysis of these 2D trimers yields valuable theoretical predictions and explanations for the inelastic neutron scattering (INS) spectra of 2D magnetic materials featuring trimerized lattices. 
\end{abstract}
\date{\today}
\maketitle
\end{CJK*}
\section{Introduction}
Elementary excitations ~\cite{pines2018elementary, RevModPhys.83.705} are key to understanding the physical properties of magnetic systems. In systems with long-range magnetic orders, the linear spin wave theory (LSWT) is commonly employed to investigate the corresponding magnetic excitations ~\cite{chumak2019magnonics, wulferding2020magnon}. For example, the LSWT without high-order corrections can match the result of inelastic neutron scattering experiments on La$_2$CuO$_4$, the parent compound of cuprate superconductors ~\cite{PhysRevLett.86.5377, PhysRevB.65.132404, peres2003spin}. Nonetheless, when quantum fluctuations are notably strong or when various types of excitations happen simultaneously, the accuracy of LSWT results is challenged, even in situations where the ground state retains magnetic order. For example, LSWT cannot adequately predict the continuum at the momentum point $(\pi,0)$ in the square-lattice Heisenberg model ~\cite{PhysRevLett.105.247001, dalla2015fractional}, which is relevant to the inelastic neutron scattering of Cu(DCOO)$_{2}$ $\cdot$ 4D$_2$O. In the theoretical analysis, the broad continuum may be due to the nearly deconfined spinon ~\cite{shao2017nearly, dalla2015fractional} or multiple-magnon scattering ~\cite{PhysRevLett.105.247001, PhysRevB.52.R15695, PhysRevLett.86.528, 10.21468/SciPostPhys.10.5.110, 10.21468/SciPostPhys.4.1.001, PhysRevLett.115.207202, 10.21468/SciPostPhys.4.1.001}. In contrast, unbiased numerical simulations are pivotal in exploring magnetic excitations in quantum many-body systems ~\cite{PhysRevLett.121.077201, PhysRevLett.118.147207, PhysRevLett.118.147206, PhysRevLett.86.528, fang2022dynamical}. Among these numerical techniques, the combination of large-scale quantum Monte Carlo with stochastic analytical continuation is a powerful technique in investigating magnetic excitations, as it can faithfully reproduce the inelastic neutron scattering spectra ~\cite{shen2019intertwined, zhou2022quantum}.

\begin{figure*}[htbp]
\centering
  \includegraphics[width=1\textwidth]{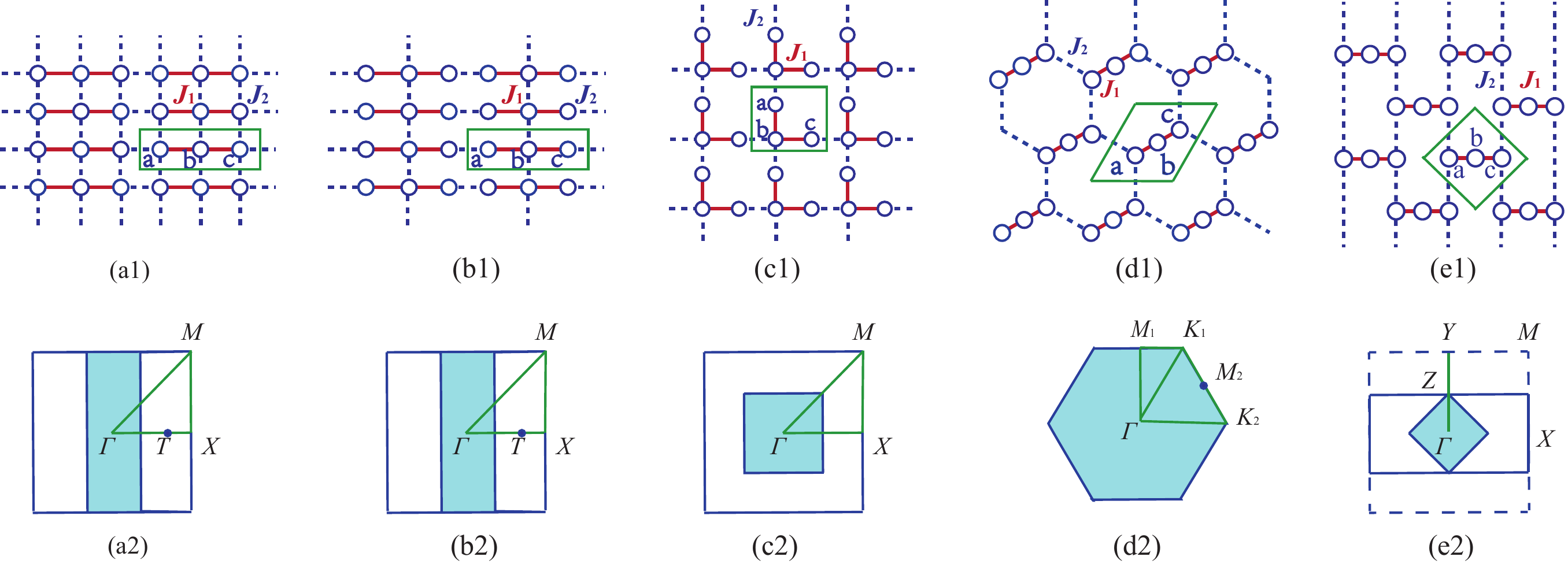}  
    \caption{Four 2D trimerized lattices and their full Brillouin zones: (a1) Collinear I lattice, corresponding to a 2D square lattice in the $g=J_2/J_1=1$ limit. (b1) Collinear II lattice, derived from the structure of CaNi$_3$(P$_2$O$_7$)$_2$ ~\cite{PhysRevB.91.104422}. (c1) Trimerized Lieb lattice exhibiting a ferrimagnetic ground state. (d1) Trimerized Hexagon lattice, derived from the structure of Ba$_4$Ir$_3$O$_{10}$ ~\cite{PhysRevLett.129.207201, cao2020quantum, cao2020quest, PhysRevB.103.224420, PhysRevB.106.075108}. (e1) a topological equivalent lattice of trimerized Hexagon lattice. The green quadrilaterals in panels (a1)-(e1) represent unit cells. The Collinear I lattice comprises three sites and six bonds within a cell, while the other lattices feature three sites and four bonds per cell. The red bonds denote the intratrimer interactions, while the dark blue ones represent the intertrimer interactions. It is worth noting that all these neighbor bonds are antiferromagnetic interactions. For (a1)-(e1), the length of red bonds is set to be one (as the length unit), and the length of blue dashed bonds in (a1)-(c1) is also equal to one, while in the (d1) and (e1), we set the length of blue bonds to be two. Besides, the lattices shown in (a1)-(c1) are also the clusters used in the CPT calculation with 24 sites (a1-b1) and 27 sites (c1). For panels (a2)-(e2), we illustrate the corresponding full Brillouin zone (BZ), and the shadow areas denote the reduced BZ. The point $T$ is $(2\pi/3,0)$, point $M_2$ is $(\sqrt{3}\pi/6,\pi/6)$, $Y$ is $(0,\pi)$ and $Z$ is $(0,\pi/2)$.}
    \label{lat}
\end{figure*}
The magnetic properties of materials contain rich physical information~\cite{jiang2016magnetic, nie2024spin}, and previous studies by some of our authors and collaborators have unveiled novel excitations in magnetically ordered systems featuring square-like lattice with $m\times n$ sublattices within unit cells~\cite{xu2019spin, yan2021magnetic, PhysRevB.98.174421, PhysRevB.99.174434, tan2023spin}, where the LSWT without high-order corrections is inadequate to describe some high-energy excitations. Further study on a trimer chain system in Ref. ~\cite{cheng2022fractional} discovered two new forms of excitations above the low-energy two-spinon continuum, these quasi-particles named ``doublon" and ``quarton", respectively. What needs to be distinguished is that the term ``doublon" also has other meanings in some literature, such as in the Hubbard and t-J models where it represents a site occupied by double holon ~\cite{PhysRevLett.104.080401,doi:10.1126/sciadv.aav2187}. In some literature, the term ``quarton" denotes flux qubits with high-order nonlinearity and long coherence times ~\cite{PhysRevLett.127.050502}. In our case, the local excitation from the ground-state doublet to the first-excited doublet with an energy $J_1$ of a trimer will be propagated by the intertrimer interaction $J_2$, forming the doublon. Meanwhile, the excitation from the ground-state doublet to the second-excited quartet leads to the formation of the quarton. Soon after, experiments on Na$_2$Cu$_3$Ge$_4$O$_{12}$ confirmed these theoretical predictions ~\cite{bera2022emergent}, where neutron experiments revealed these two excitations. Moreover, introducing an additional magnetic field to the trimer chain system induces $XY$ phase and $1/3$ magnetization plateau phase, and the doublon and quarton are still observable when $g=J_2/J_1$ and magnetic field are weak~\cite{cheng2024quantum}. 

In two-dimensional (2D) systems, some magnetic materials also have trimerized structures, as evidenced in some experimental studies, including compounds like CaNi$_3$(P$_2$O$_7$)$_2$, Ba$_4$Ir$_3$O$_{10}$, and Ba$_4$Ru$_3$O$_{10}$ ~\cite{PhysRevB.91.104422, PhysRevLett.129.207201, PhysRevB.84.054439, igarashi2015effects}. Unlike the 1D chain, 2D trimer systems exhibit magnetic ground states with magnons as their low-energy excitations. When the intertrimer interaction (or $g=J_2/J_1$) are relatively weak, doublon and quarton can be expected. Therefore, exploring how magnon, doublon, and quarton evolve as the ratio $g$ changes is of great interest. 

There are plenty of 2D trimerized models, such as the trilayer models ~\cite{10.21468/SciPostPhys.12.2.054, yang2022equipartition}, the Kagome lattice ~\cite{PhysRevB.99.165141, farnell2019emergence, chen2012quantum, farnell2019emergence} and three-leg ladder ~\cite{PhysRevB.103.125120}. Certainly, it is impossible to
study all the 2D trimer structures. In our study, we focus on four lattices, including the Collinear I, Collinear II, trimerized Lieb lattice, and trimerized Hexagon lattice, which are shown in Figs.~\ref{lat}(a1)-(d1). The Collinear I lattice, as quite a natural generalization of the 1D trimer model, corresponds to a 2D square lattice in the $g$=1 limit. Using this lattice, we can explore how magnon, doublon, and quarton emerge into one magnon mode from weak $g$ to $g$=1. For the Collinear II lattice and trimerized Hexagon lattice, some materials exhibit these lattice structures such as CaNi$_3$(P$_2$O$_7$)$_2$ ~\cite{PhysRevB.91.104422} and Ba$_4$Ir$_3$O$_{10}$ ~\cite{PhysRevLett.129.207201, cao2020quantum, cao2020quest, PhysRevB.103.224420, PhysRevB.106.075108}. As for the choice of trimerized Lieb lattice, its ground state is a ferrimagnetic order, which is very different from the antiferromagnetic orders of some other lattices. It would be interesting to explore the evolution of the magnon, doublon, and quarton in a ferrimagnetic ground state system. For trimerized systems sharing the same local trimers, the doublon and quarton always appear in the weak intertrimer interaction region. Still, their dispersion relations and evolutions with $g$ depend on lattice geometries. These models all share a common characteristic: they feature intratrimer exchange interaction represented by $J_1$ and intertrimer exchange interaction labeled as $J_2$. All of these lattices consist of trimer blocks with $a$, $b$, and $c$ sublattices, as depicted in Figs.~\ref{lat}(a1)-(d1). A slight distinction among these lattices is that the last exhibits a hexagon structure, while the others are arranged like a square or rectangle lattice, however, we can transform the hexagon structure to a topologically equivalent square-like lattice, as shown in Fig.~\ref{lat}(e1).

\begin{figure}[htbp]
\centering
  \includegraphics[width=0.48\textwidth]{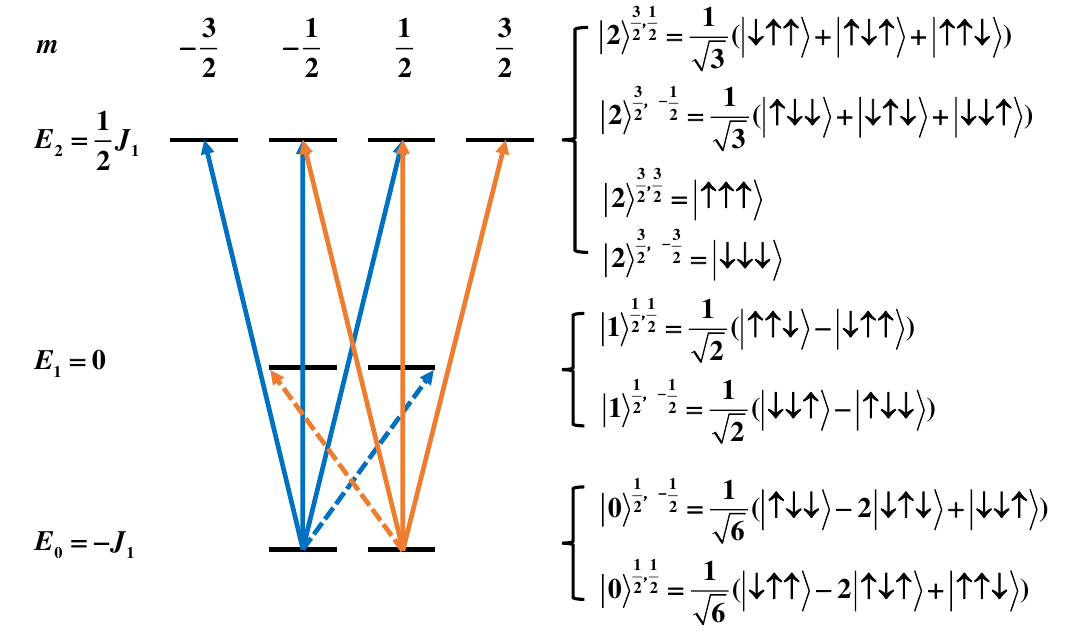}
    \caption{The energy levels and the corresponding wave functions of an isolated trimer block. Its ground state is a doublet with effective spin-$1/2$. $\left| n \right\rangle ^{S,m}$ denotes an eigenstate, where $n=0, 1, 2$ represents the ground state, the first-excited state, and the second-excited state, respectively. $S$ is the total spin quantum number, and $m$ is the magnetic quantum number. The dashed lines denote the $|\Delta m|=1$ excitations, and the solid lines are $|\Delta S|=1$ excitations. The orange lines denote the excitations from $\left| 0 \right\rangle ^{\frac{1}{2},\frac{1}{2}}$ ground state and the blue lines are the excitations from $\left| 0 \right\rangle ^{\frac{1}{2},-\frac{1}{2}}$ ground state.}
    \label{late}
\end{figure}
This paper studies the dynamic spin structure factors, denoted as $S(q,\omega)$, for four trimerized Heisenberg models with varying ratios $g=J_2/J_1$. To calculate this quantity, we use the stochastic series expansion quantum Monte Carlo (SSE QMC) combined with the stochastic analytic continuation (SAC) method ~\cite{gull1984maximum, PhysRevE.94.023303, PhysRevB.57.10287, PhysRevE.94.063308, beach2004identifying}, which has gained a lot of improvement in recent years and has been successfully applied to many systems. Meanwhile, we also use the cluster perturbation theory (CPT) with exact diagonalization (ED) as a solver to supplement QMC results. In the 2D trimer models, unlike the 1D trimer chain that exhibits  two-spinon continuum behaviour, the low-energy part is dominated by magnons. For these low-energy magnons, we employ perturbative analysis to derive the effective models among block spins formed by the trimer doublets ~\cite{lowdin1951note}. Subsequently, we can apply linear spin wave theory (LSWT) to the effective models and derive the dispersions of low-energy magnons ~\cite{PhysRevLett.97.207202}. For the doublon and quarton, we use perturbative analysis (PA) to extract their optimal dispersion relations ~\cite{kato2013perturbation}, which is a suitable approach to study the dispersion relation of the $S(q,\omega)$ spectra for weak $g$. However, as $g$ gradually increases, the doublons and quartons merge into magnon modes or become continua. The Collinear II and the trimerized Hexagon lattice maintain more features inherited from the trimer chain, resulting in a broad, high-energy continuum. Through a combination of numerical simulations and theoretical analyses, our research comprehensively studies the excitation dynamics of 2D trimer models. Our results provide a better understanding of the excitation mechanisms in 2D trimer block systems and the corresponding materials.

We have structured this paper as follows. In Sec.~\ref{Models and Methods}, we introduce four 2D trimerized models featuring antiferromagnetic interactions and a brief overview of the numerical techniques employed. Moving on to Sec.~\ref{Numerical Results}, we present the spectra of four trimerized systems and draw comparisons between numerical data obtained from several methods. In Sec.~\ref{Analysis}, we focus on analyzing quasi-particles at various energy levels, including magnons, doublons, and quartons. Finally, in Sec.~\ref{Summary and Discussion}, we summarize the discussion of our studies and some of our plans for the future. 

\section{Models and Methods} \label{Models and Methods}
\subsection{Model Hamiltonian}
\label{ha}
In this paper, we thoroughly explore the isotropic Heisenberg model on four distinct 2D trimerized lattices, including the Collinear I, Collinear II, trimerized Lieb lattice, and trimerized Hexagon lattice, illustrated in Figs.~\ref{lat}(a1)-(d1). Each lattice features two kinds of nearest-neighbor exchange interactions, denoted as $J_1$ and $J_2$. The Hamiltonian for these models is given by 
\begin{equation}\label{eq1}
H = {J_1}\sum\limits_{\left\langle {i,j} \right\rangle } {{S_i} \cdot } {S_j} + {J_2}\sum\limits_{\left\langle {i,j} \right\rangle^ \prime } {{S_i} \cdot } {S_j},
\end{equation}
where $S_i$ represents the spin operator at site $i$, $J_1$ and $J_2$ represent the intratrimer and intertrimer coupling strengths, respectively. The first term (inclusive of $J_1$) is also denoted as $H_t$, representing interactions within a trimer, while the second term (inclusive of $J_2$) is denoted as $H_{tt}$, indicating interactions between two trimers. For simplicity, we define the coupling ratio as $g=J_2/J_1$,  where $g=0$ corresponds to the decoupled trimer limit and $g=1$ represents the uniform cases. We set $J_1=1$ as the energy unit and let $g$ vary in the range of $(0,1]$ to explore the dynamic spin structure factors. The corresponding full BZs are shown in panels Figs.~\ref{lat}(a2)-(d2), and the reduced BZs are illustrated in shadow areas.

This paper mainly focuses on the magnetic excitations of these trimerized lattice models. The magnetic excitation spectra with $\Delta S=1$ can be well revealed by the dynamic spin structure factors $S(q,\omega)$, and can be detected by the inelastic neutron scattering experiment. To calculate this quantity, we mainly use two kinds of methods: QMC and CPT. Each method has its advantages and disadvantages. In addition, we also used LSWT to analyse the low-energy magnon and perturbative analysis to determine the dispersions of doublon and quarton. In the following, we mainly introduce the QMC and CPT methods.

\subsection{Quantum Monte Carlo}
\label{qmc}
QMC can be simulated using the large-scale lattice, thereby capturing more information on long-range correlation and entanglement ~\cite{PhysRevB.106.085101}. However, this method does not provide direct access to real-time or real-frequency dynamic correlation functions. Instead, these quantities are obtained through analytical continuation from imaginary-time Green functions. However, with the development of powerful stochastic analytical continuation methods ~\cite{gull1984maximum, PhysRevE.94.023303, PhysRevB.57.10287, PhysRevE.94.063308, beach2004identifying, PhysRevB.76.035115, reichman2009analytic, PhysRevB.78.174429, PhysRevE.81.056701, PhysRevB.101.085111, PhysRevB.102.035114}, researchers have continually improved the computational ability to get high-resolution excitation spectra. 

To obtain the dynamic spin structure factors using QMC, we start by obtaining the imaginary-time correlation functions through SSE-QMC samplings. These correlation functions are defined as ${G_q}(\tau)= 3\left\langle S_{ - q}^z(\tau )S_q^z(0)\right\rangle$, where the factor of $3$ arises from the $SU(2)$ continuous symmetry of isotropic Heisenberg model, and $S_{q}^z$ represents the Fourier transform of the $z$-component spin operator, which can be written as:
\begin{equation}
    S_q^z = \frac{1}{{\sqrt N }}\sum\limits_{i = 1}^N {{e^{ - i{{\vec r}_i} \cdot \vec q}}S_i^z},
\end{equation}
where the atomic coordinates ${\vec r}_i$ is taken in the Fourier transform. So, we use the so-called ``atomic gauge". To visualize the $S(q,\omega)$, we follow a high-symmetry path $\Gamma (0,0)$ $\rightarrow$ $T(2\pi/3,0)$ $\rightarrow$ $X(\pi,0)$ $\rightarrow$ $M(\pi,\pi)$ $\rightarrow$ $\Gamma (0,0)$ for three square-lattice-like models, as depicted in Figs.~\ref{lat}(a1)-(c1). We use the full BZ for these lattices when doing the Fourier transform. However, for the trimerized Hexagon lattice, we employ the dynamic spin structure factors in the lattice reduced BZ, which is shown in Fig. ~\ref{lat}(d2), and the reduced $S_{red}^{zz}(q,\omega)$ can be written as:
\begin{equation}
    S_{red}^{zz}(q,\omega) = \sum\limits_{\alpha = a,b,c} {S_{\alpha \alpha}^{zz}(q,\omega)},
\end{equation}
where $S_{\alpha \alpha}^{zz}(q,\omega)$ is the dynamic structure factors which only consider the spins at sites $\alpha \in {a, b, c}$, and we choose the path $\Gamma (0,0)$ $\rightarrow$ $M_1 (0,\pi/3)$ $\rightarrow$ $K_1 (\sqrt{3}\pi/9,\pi/3)$ $\rightarrow$ $\Gamma (0,0)$ $\rightarrow$ $K_2 (2\sqrt{3}\pi/9,0)$ $\rightarrow$ $M_2 (\sqrt{3}\pi/6,\pi/6)$ $\rightarrow$ $K_1 (\sqrt{3}\pi/9,\pi/3)$. The connection between ${G_q}(\tau )$ and the dynamic spin structure factors $S(q,\omega)$ is established through analytical continuation, which is expressed by the following equation,
\begin{equation}
    {G_q}(\tau ) = \frac{1}{\pi }\int_{ - \infty }^\infty  {d\omega S(q,\omega ){e^{ - \tau \omega }}}.
\end{equation}

To tackle this inverse problem, in the SAC procedure, we parameterize $S(q,\omega)$ by employing multiple $\delta$ functions and adjust both their amplitudes and frequencies when sampling the spectrum with the probability distribution:
\begin{equation}
    P(S) \propto exp(-\frac{\chi^2}{2\Theta}),
\end{equation}
where $\chi^2$ is the standard goodness of fit and $\Theta$ is a fictitious sampling temperature that can be adjusted. Stochastic averaging of the configurations balancing $\chi^2$ minimization and sampling entropy provides converged results of the spectral functions ~\cite{shao2023progress}. 

In this work, our calculations for the dynamic spin structure factors are conducted on $L_x \times L_y$ trimer blocks, with $L_y=3L_x=48$ for the two collinear models and $L_y=L_x=24$ for the trimerized Lieb lattice and trimerized Hexagon lattice. These systems have periodic boundary conditions. Besides, we set the inverse temperature to $\beta=256$, allowing us to access excitation modes at very low energy scales.

\subsection{Cluster Perturbation Theory}
\label{cpt}
Cluster perturbation theory (CPT) offers an alternative method for obtaining $S(q,\omega)$ without analytical continuation. CPT employs exact diagonalization (ED) as a solver and has been successfully applied in the study of several spin models ~\cite{yu2018deconfinement, senechal2000spectral, ovchinnikov2010cluster}. In this approach, ED is employed to investigate the dynamic correlation effects within the clusters. At the same time, perturbation and mean-field theories are employed to deal with the intercluster interactions and correlations. Within the CPT framework, we focus on calculating the transverse dynamic spin structure denoted as ${S^{ +  - }}(q,\omega)$. In 2D trimer models with $SU(2)$ symmetry, ${S^{ +  - }}(q,\omega )$ are similar to the $S^{zz}(q,\omega )$. 

The cluster size is limited for the CPT method with ED as a solver. We use the clusters with $n_x\times n_y\times 3$ sites, where 3 comes from the number of sublattices in one unit cell. For two collinear trimerized models, we set $n_x=2$ and $n_y=4$. We set $n_x=3$ and $n_y=3$ for the trimerized Lieb lattice, as shown in Figs. ~\ref{lat}(a1)-(d1). Both local and non-local dynamic correlations are treated exactly within clusters. However, non-local correlations between clusters are addressed using perturbation and mean-field theory. It is worth noting that the mean-field treatment tends to overestimate magnetic ordering ~\cite{PhysRevB.93.075131, PhysRevB.70.245110}, implying a potential underestimation of quantum fluctuations. Nevertheless, this method yields nearly exact results within certain limits. The first limit is that the cluster is infinitely large. This allows for the convergence results with increasing cluster sizes, which is more efficient than calculations with finite-size torus geometries ~\cite{RevModPhys.77.1027}. Another exact limit occurs when the interactions between clusters are infinitely small, making perturbation theory work well. Therefore, the CPT method is very suitable to analyze the excitations in weak $g$. 

\section{ Numerical Results} \label{Numerical Results}
When $g=0$, all four models are fully decoupled, and each trimer block forms a doublet ground state with an effective spin $S=1/2$. The energy levels of a single trimer have been previously depicted in Ref ~\cite{cheng2022fractional}, which are also shown diagrammatically in Fig.~\ref{late}. The ground state and the first-excited state are doublets with the magnetic quantum number $m=\pm 1/2$, and the second excited state is a quartet with $m=\pm 1/2$ and $\pm 3/2$. Fig.~\ref{late} also illustrates the $|\Delta S|=1$ (quarton excitations, solid arrow lines) and $|\Delta m|=1$ ($|\Delta S|=0$, doublon excitations, dashed arrow lines) excitations of a single trimer, which can be used to explore the dispersions of quasiparticles using perturbative analysis. As we turn to small values of $g$, the low-energy excitations are dominated by the effective block spin models, which will be analyzed in detail in Sec.~\ref{Analysis}. 
\begin{figure}[htbp]
  \centering
  \includegraphics[width=0.48\textwidth]{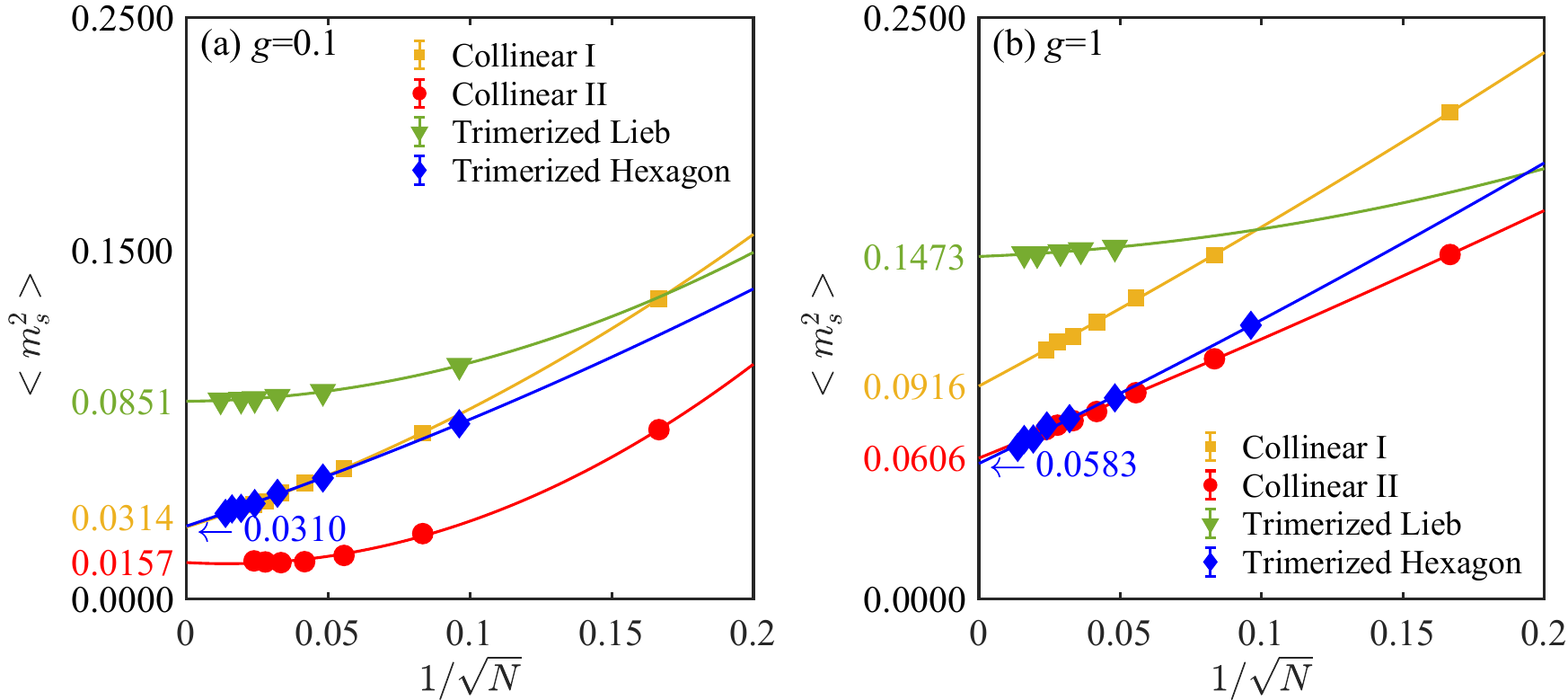}
    \caption{The extrapolations of square staggered magnetization on four lattices with (a) $g$=0.1 and (b) $g$=1. Error bars are much smaller than the size of the symbols. The extrapolated values are shown directly on the horizontal axis. To avoid text overlap, we represented the intercepts of the trimerized Hexagon using arrows.}
    \label{ms2}
\end{figure}

Specifically, the introduction of $J_2$ enables the excitation from the one-trimer ground-state doublet to the first-excited doublet state, to propagate as a quasi-particle moving on the lattices. We refer to this quasi-particle as a doublon. Similarly, introducing $J_2$ also induces the quasiparticle named quarton to propagate in the 2D lattices for the one doublet to quartet excitation. The doublon and quarton have their typical energy scales for spin-$1/2$ systems, $J_1$ and $1.5J_1$ respectively. As we vary the value of $g$, how the low-energy magnon, intermediate-energy doublon, and high-energy quarton evolve is much less known for the 2D cases. Under weak $g$, the separation of doublon and quarton can be observed in the CPT spectrum. To further understand the spectra of doublon and quarton, we numerically analyze hundreds of dispersions obtained from perturbative analysis (PA) and find some optimal dispersions that can match the CPT dispersions (the peak positions of $S^{+-}(q,\omega)$ at certain $\omega$ along $q$) quite well. The following sections will give detailed analyses and discussions of excitations in terms of the dynamic spin structure factors.

It is worth noting that when $g$ falls in the $(0,1]$, the ground states of four models exhibit long-range magnetic orders. To assess and compare the relative strengths of magnetic orders in various models, we representatively choose $g=0.1$ and $g=1$ to perform finite-size extrapolations. The order parameter employed in our analysis is the square staggered magnetization, denoted as $m_s^2$,
\begin{equation}
    {m_s^2}  = \frac{3}{{{N^2}}}{\left| {\sum\limits_i {{\phi _i}{S_i^z}} } \right|^2},
\end{equation}
where $\phi_i=\pm 1$ represents staggered phase factors, and the factor of 3 comes from the isotropic strength of all three spin components. To perform the extrapolations, we employ second-order polynomial fittings, the results are presented in Fig.~\ref{ms2}. Notably, in the case of $g=1$, where the Collinear I lattice is equivalent to the square lattice, our extrapolated value for $m_s^2$ aligns well with the results in Ref ~\cite{sandvik2010loop, Gerber_2009, PhysRevLett.80.1742}. In addition, the trimerized Hexagon lattice shares the same diluted configuration as the Collinear II lattice, resulting in nearly the same extrapolated values, as illustrated in Fig.~\ref{ms2}(b). The extrapolated values are shown directly in the horizontal axis of Fig.~\ref{ms2} using corresponding colors.
\begin{figure*}[htbp]
    \centering
    \includegraphics[width=1\textwidth]{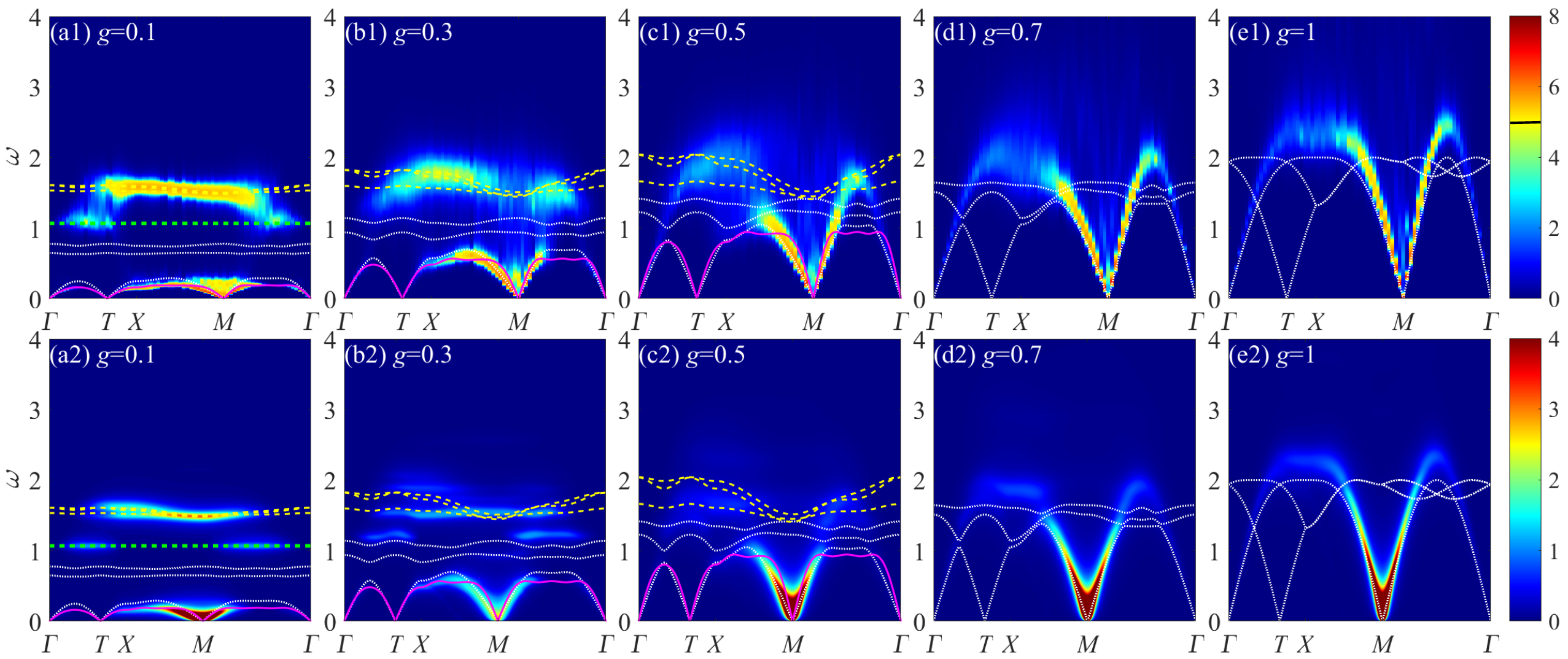}
    \caption{The dynamic spin structure factors of the Collinear I lattice, obtained through QMC and CPT methods, are presented with varying ratios $g=J_2/J_1$. We follow a high-symmetry path $\Gamma (0,0)$ $\rightarrow$ $T(2\pi/3,0)$ $\rightarrow$ $X(\pi,0)$ $\rightarrow$ $M(\pi,\pi)$ $\rightarrow$ $\Gamma(0,0)$, as shown in Fig. ~\ref{lat}(a2), to show the dynamic spin structure factors. The solid lines are the LSWT results of its low-energy effective block spin model. In contrast, the white dotted lines correspond to the LSWT results of the original spin model. The dashed lines in the higher-energy parts denote the optimal dispersions obtained by the PA. Panels (a1)-(e1) display results from QMC-SAC, while panels (a2)-(e2) feature CPT results. For clarity, in the QMC-SAC results, a piecewise function is utilized where the intensity is bifurcated at $U_0=5$. When the value is less than 5, the low-intensity section follows a linear distribution. Beyond this threshold, a logarithmic scale is applied, expressed as $U = {U_0} + {\log _{10}}{S(q,\omega )}-{\log _{10}}{U_0}$. However, no additional transformation is applied to the CPT spectrum. This same piecewise function is also used in the subsequent excitation spectra of other lattices.}
    \label{coex}
\end{figure*}
\subsection{Collinear I Lattice}
\label{co1}
The first two-dimensional (2D) expansion of trimer blocks we study, termed Collinear I, is depicted in Fig.~\ref{lat}(a1). This structure is a simple expansion of the 1D trimer chain. A main feature of Collinear I is that it evolves into the uniform square-lattice Heisenberg model when the ratio $g=J_2/J_1$ is set to be 1. Fig.~\ref{coex} shows the dynamic spin structure factors $S(q,\omega )$ changing with $g$. Figs.~\ref{coex}(a1)-(e1) represent QMC-SAC results for various $g$, while Figs.~\ref{coex}(a2)-(e2) display CPT results, corresponding to the same $g$ as used in the QMC-SAC data. This arrangement allows for a direct comparison between two methods under the same $g$ ratio.

From the QMC-SAC data, at $g=0.1$, we notice that the doublon band is approximately at $\omega \approx J_1$, and the quarton band is near $\omega \approx 1.5J_1$. When $g=0.3$, these two spectra seem to start melting into each other and then they are highly mixed. For the clearer separation of higher bands, we have increased the Monte Carlo samples and let the error of $S(q,\omega )$ below $10^{-4}$. This approach provides more reliable SAC results. We have more discussions about the performance of SAC calculation in Appendix ~\ref{SAC-test}. As $g$ further increases, the higher part of the excitation spectrum begins to exhibit a continuum near point $X$ in Fig. ~\ref{coex}(e1), and this continuum persists up to $g=1$ which is in the square-lattice Heisenberg limit. However, the 24-site cluster used in the CPT is still not large enough to fully capture this continuum in Fig. ~\ref{coex}(e2). In Ref. ~\cite{yu2018deconfinement}, they have already identified that the continuum only has a very small proportion in CPT results. Due to the antiferromagnetic ground state, the magnon band has gapless Goldstone mode at $M$ and $\Gamma$ points. The minimal spectral weight at the $\Gamma$ point is due to the conservation of total $S^z$. With an increase in $g$, the magnon bandwidth and spin-wave velocity rise. This is accompanied by the elevation of the low-energy magnon spectrum near the point $X$ and the diminishing in spectral weights. When $g=0.7$, the low-energy magnon has already merged with the higher-energy part, forming a single magnon mode accompanied by some continuum.

The CPT results exhibit similar behavior in the low-energy magnon band. However, the doublon and quarton bands are well separated at the weak $g$ ratios. Regarding the doublon, as the ratio $g$ increases, it becomes more dispersive and eventually forms a dome, becoming a significant part of the single-magnon mode. For the quarton, there appear to be energy splittings with a diminishing spectral weight in the high-energy part and a broadening spectrum near $\omega \approx 1.5J_1$. The doublon and quarton eventually merge into the low-energy mode when $g$ approaches 1. 

The comparison between QMC and CPT in Fig.~\ref{coex} reveals notable differences in the doublon and quarton bands. The primary distinction lies in the capabilities of each method, as highlighted in Sec.~\ref{cpt}. The CPT method is exact in the $g\rightarrow 0$ limit. Coupled with the high-accuracy real-frequency dynamics of the ED solver, it provides detailed information on the dispersions of magnetic quasiparticles for weak $g$. In this case, PA also works effectively (the details of PA are shown in Sec.~\ref{magnon}). The dispersions obtained from PA match the quarton band of CPT results quite well at weak $g$, as shown in the yellow dashed lines of Fig.~\ref{coex}. The most relevant dispersions are presented in Table ~\ref{tab} of Sec.~\ref{doqu}. However, for larger $g$, the CPT method as a cluster mean-field treatment of magnetic order, may tend to overestimate the magnetic order. Therefore, when the cluster is not large enough, capturing the high-energy broad continuum becomes challenging. In contrast, QMC with large-scale simulation allows us to obtain a confident spectrum from high-quality imaginary-time Green functions using stochastic analytical continuation (SAC), as demonstrated in Ref.~\cite{shao2017nearly}. Regarding the not-so-well separation of doublon and quarton of SAC data at $g=0.1$, PA in Sec.~\ref{doqu} also indicates some overlap between the doublon and quarton bands. It would be harder for SAC to get a clear separation of these two bands with insufficient accuracy of $S(q,\tau)$. We have added more discussion in Appendix~\ref{SAC-test}.

\begin{figure*}[htbp]
     \centering
    \includegraphics[width=1\textwidth]{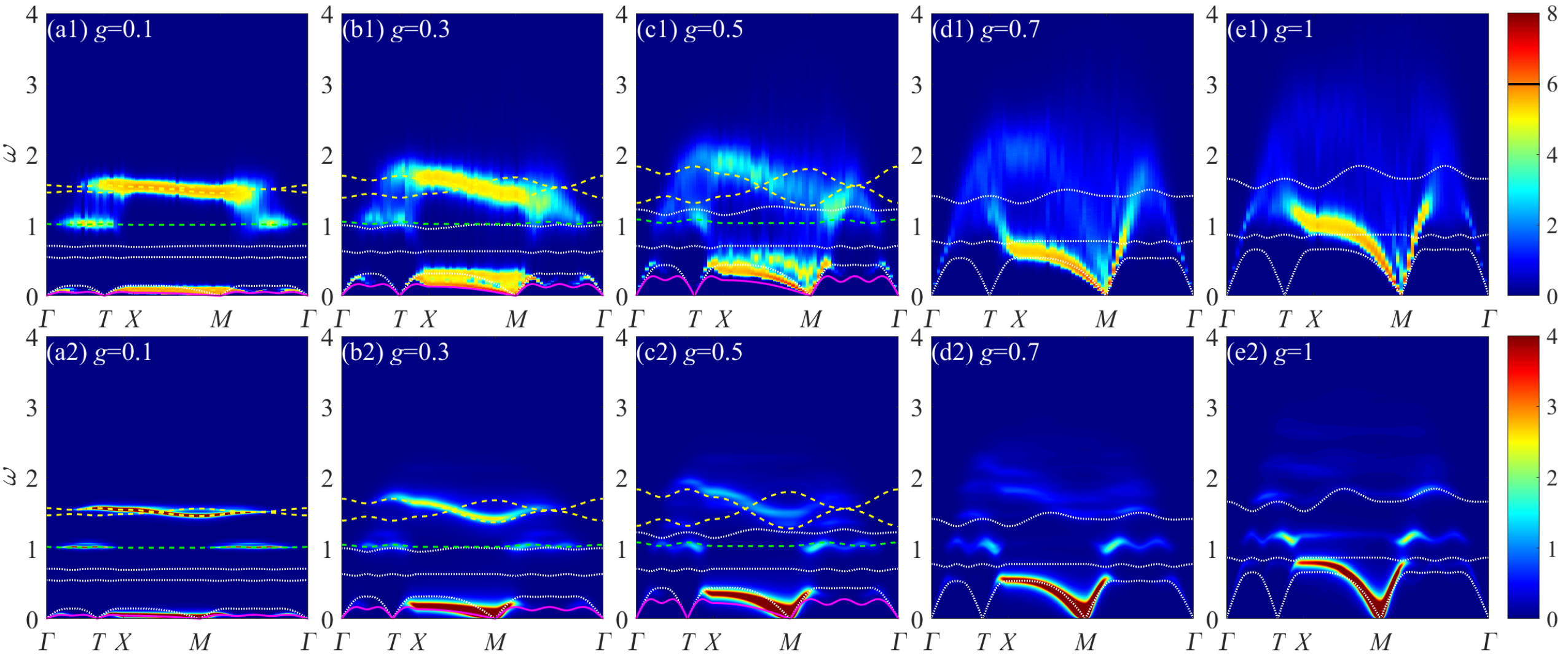} 
     \caption{Dynamic spin structure factors of the Collinear II lattice for different $g$ values. The pink solid lines represent the magnon dispersion of the corresponding effective model formed by the trimer block spins, while the white dotted lines illustrate the LSWT results for the original model. The yellow and green dashed lines represent the dispersions of quarton and doublon, respectively, obtained from the PA in weak $g$. Panels (a1)-(e1) show results obtained from the SAC method, and (a2)-(e2) are the results obtained from the CPT. The high-symmetry path is the same as the Collinear I model as shown in Fig.~\ref{lat}(b2), and we use the same logarithmic scale in SAC results when $S(q,\omega)>U_0=5$, expressed as $U = {U_0} + {\log _{10}}{S(q,\omega )}-{\log _{10}}{U_0}$.}
     \label{j3ex}
\end{figure*}
As the ground state resides in the N\'eel phase, we employ the LSWT to further study the low-energy magnon using two models. The first one is the original spin model on the Collinear I lattice with N\'eel antiferromagnetic order. Another one is the effective spin-$\frac{1}{2}$ antiferromagnetic model on a rectangle lattice formed by the block spin of each trimer. The effective model has vertical exchange coupling $J_v\approx g$ and horizontal coupling $J_h \approx \frac{4}{9}g$, as shown in Fig.~\ref{eff}(a), more details about how to derive the effective interactions can be seen in Sec.~\ref{magnon}. In Fig.~\ref{coex}, the white dotted lines illustrate the LSWT results of the original model, while the pink solid line shows the LSWT results of the effective model. Notably, there is a gapless point at $T (\frac{2}{3}\pi,0)$, which is due to the BZ folding. When $g$ is weak, as depicted in Figs.~\ref{coex}(a1)(a2), both the spin wave velocity and the curve of white dotted lines deviate from the magnon band of QMC and CPT. However, the pink solid lines presenting LSWT results of the effective model match the magnon band well, with better matching as $g$ decreases. In contrast, as $g$ becomes larger, the effective model becomes more inadequate to capture the low-energy physics. Therefore, we only present LSWT results of the effective model under $g\le 0.5$. Due to the stronger ground-state magnetic order, the LSWT results of the original model match the magnon band increasingly better. At $g=1$, the linear spin wave theory can match the single magnon mode quite well. 

\subsection{Collinear II Lattice }
\label{co2}

The second trimer model we study is the Collinear II lattice which can be found in the CaNi$_3$(P$_2$O$_7$)$_2$ magnetic material~\cite{PhysRevB.91.104422}. Although the nickel ion has an effective spin quantum number of $S=1$ ~\cite{PhysRevB.93.184409, PhysRevB.97.224413, PhysRevB.74.024430}, our study only focuses on the $S=1/2$ scenario. The Collinear II lattice, shown in Fig.~\ref{lat}(b1), includes three sites and four connected bonds per unit cell. Compared with Collinear I, there is the dilution of vertical bonds along the $y$ direction, and a broad continuum observed at larger $g$ values, a characteristic of 1D chain features, which can be explained as the dimensional reduction effect ~\cite{PhysRevLett.126.227201}, more numerical identifications can be found in the Appendix ~\ref{dimensional}. The dynamic spin structure for various $g$ values, alongside effective dispersions and LSWT results, are shown in Fig.~\ref{j3ex}. The BZ for Collinear II is identical to that of Collinear I, so the path followed. Similarly, the ground state of this model also resides in the N\'eel phase. 

In Figs.~\ref{j3ex}(a1)-(e1), the dynamic spin structure factors $S(q,\omega)$ of Collinear II exhibit similarities to Collinear I. Notably, the points at $\Gamma$, $M$ and $T$ are gapless. For $g=0.1$, as seen in Fig.\ref{j3ex}(a1), the excitation spectrum distinctly separates into magnon, doublon, and quarton parts. The magnon has a gap to the doublon at $g=0.1$. The doublon around $\omega \approx J_1$, and the quarton near $\omega \approx 1.5J_1$. In Fig.\ref{j3ex}(a1), the doublon and quarton are more distinctly separate even in the SAC data due to more localized excitations with fewer connection bonds along the $y$ direction compared to Collinear I. However, along the path $\Gamma-T$ and $M-\Gamma$, the strong fluctuation in the high-energy parts among adjacent momenta again suggests a limitation of SAC performance due to the barely good enough QMC data qualities, as evidenced by further discussions in Appendix ~\ref{SAC-test}.

As $g$ increases, the few connection bonds along the $y$ direction introduce quasi-1D physics into the dynamic spin structure factors. The non-uniformity of the correlation effect caused by the exchange interaction in the $x$ and $y$ directions becomes increasingly pronounced in the SAC spectrum. In this case, the doublon and quarton mix and melt into each other and form a broad continuum, while the energy band of the low-energy magnon, continues to expand upwards with the vanishing of the spectrum around point $T$ and eventually becomes the lower bound of the continuum spectrum.

\begin{figure*}[htbp]
    \centering
    \includegraphics[width=1\textwidth]{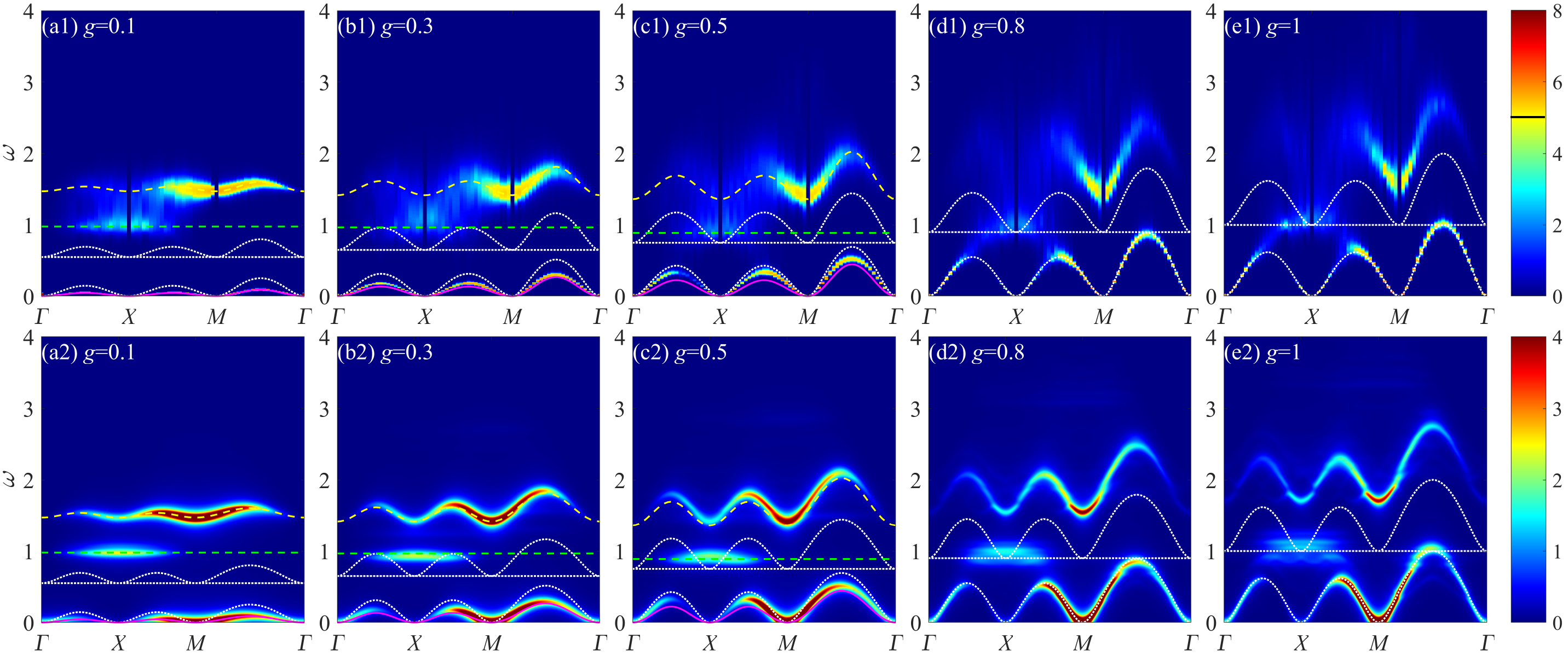}
    \caption{Dynamic spin structure factors of the trimerized Lieb lattice at various $g$ values. The pink solid lines represent the magnon dispersions of the corresponding effective model formed by the trimer block spins. The white dotted lines illustrate the LSWT results of the original model. The yellow and green dashed lines represent the dispersions of quarton and doublon, respectively, obtained from the PA in weak $g$. Panels (a1)-(e1) show results obtained from the QMC-SAC method, while (a2)-(e2) present results obtained from CPT. The high-symmetry path is shown in Fig.~\ref{lat}(a2), and we use the same logarithmic scale in QMC-SAC results when $S(q,\omega)>U_0=5$, expressed as $U = {U_0} + {\log _{10}}{S(q,\omega )}-{\log _{10}}{U_0}$.}
    \label{liebex}
\end{figure*}
For the CPT results shown in Figs.~\ref{j3ex}(a2)-(e2), the dispersions of doublon and quarton are revealed in detail when $g\le 0.3$. The CPT data is used to obtain the optimal curve from PA (see Sec.~\ref{doqu}), shown with yellow dashed lines for quarton and green dashed lines for doublon. The details of the dispersions are shown in Table.\ref{tab} of Sec.~\ref{doqu}. When $g=0.3$, the spectrum weight of one quarton dispersion vanishes in the excitation spectrum, as shown in Fig.~\ref{j3ex}(b2). Due to the limitations of the small cluster used in CPT and the overestimation of magnetic order, the high-energy broad continuum cannot be well characterized. However, the low-energy magnon band closely resembles the QMC results. In Figs.~\ref{j3ex}(a2)-(e2), we set the Lorentz broadening factor $\eta=0.01$ for $g=0.1$ and $\eta=0.05$ for other $g$ values. 

To analyze the low-energy magnon, we also calculate the LSWT results of the effective model and the original model, shown with dotted white lines and pink solid lines in Fig.~\ref{j3ex}, respectively. The smaller $g$ is, the better the LSWT of the effective model can match. The effective model is an antiferromagnetic Heisenberg model on a rectangle lattice with different vertical exchange interaction $J_v \approx g/9$ and horizontal interaction $J_h \approx 4g/9$, as can be seen in Fig.~\ref{eff}(b) and Sec.~\ref{magnon} with more details. When $g=0.1$, the pink solid line aligns closely with the QMC-SAC and CPT results. Particularly in the path from $M$ to $\Gamma$, as depicted in Fig.~\ref{j3ex}(a1), the pink line accurately matches the ripple-like dispersions observed in the low-energy magnon excitation spectra. However, it is observed that the effectiveness of this effective model diminishes with an increase of $g$. For instance, at $g=0.5$, the pink solid line deviates significantly from the magnon part of QMC-SAC and CPT spectra. This deviation becomes even more pronounced at larger $g$. In contrast, as $g$ increases, the spin wave velocity of the lowest branch obtained by LSWT on the original model tends to align with the QMC one, although the total dispersion band requires high-order corrections of spin wave theory. 

\subsection{Trimerized Lieb Lattice}
\label{lieb}
The trimerized Lieb lattice depicted in Fig.~\ref{lat}(c1) is formed by folding the trimer into a perpendicular block comprising three sites and four bonds. This lattice corresponds to a $1/4$-depleted square lattice when $g=1$. The dynamic spin structure factors for this lattice model are presented in Fig.~\ref{liebex}. It follows the same BZ path as that of Collinear I and Collinear II. We must note that the gapless points in $X$ and $M$ need a very large $\beta$ to obtain convergence results; thus, we didn't show them in Fig.~\ref{liebex}. However, we can infer its behavior from the points around it.

In Figs.~\ref{liebex}(a1)-(e1), we can discern three distinct excitation bands in the QMC calculations. A notable feature is a broad band with very weak dispersion, at around $\omega \approx J_1$. As the $g$ increases, this band's energy slightly decreases and then increases back to around $\omega\approx J_1$. The origin of the flat band in trimerized Lieb lattice is similar to the other ferromagnetic lattices, as a consequence of destructive interference between different ``hopping" paths of quasiparticles like doublon ~\cite{PhysRevLett.121.096401, PhysRevLett.99.070401, yin2019negative, li2021dirac}. Concurrently, the bandwidths of the low-energy magnon and high-energy quarton expand. In particular, the spectral associated with the doublon and quarton around point $X$ becomes broad, indicating the possible strong scattering between these two quasi-particles. When $g=1$, the upper boundary of the magnon aligns with $\omega=J_1$ at the middle point of $M$ to $\Gamma$, and the low-energy magnon band closely matches the LSWT prediction on the original model (see white dotted line in Fig.~\ref{liebex} (e1)) without any further correction. This is due to the ferrimagnetic ground state with effective ferromagnetic exchange interaction between trimers, reflected in the quadratic dispersion at around $\Gamma$, $X$, and $M$. However, the broad continuum and the possible separation of the upper two bands cannot be described by the LSWT of the original model. In addition, when $g$ is weak, like $g=0.1$, the LSWT of the effective block spin model with the same ferromagnetic exchange interaction along $x$ and $y$ directions is more suitable to match the dispersion of low-energy magnon. The effective interaction strength is $-2g/9$ from our calculation using the Kadanoff method in Sec.~\ref{magnon}. 

For the CPT results shown in Figs.~\ref{liebex}(a2)-(e2), we always observe three separate excitation bands. The CPT results are more reliable in the weak $g$ regime, as mentioned in Sec.~\ref{cpt}. We can further use PA to quantitatively study the doublon and quarton bands of CPT results. More details can be found in Sec.~\ref{doqu}. These optimal curves (yellow dashed lines for quartons and green dashed lines for doublons) can also match the QMC data quite well. Especially for quarton, the dispersion obtained from PA can match the high-energy band quite well even at $g=0.5$. However, due to the limitation of the small cluster, when $g$ gets larger, CPT may overestimate the magnetic order or underestimate the quantum fluctuation, leading to the failure in characterizing the high-energy broad continuum around point $X$.
\begin{figure}[htbp]
    \centering
    \includegraphics[width=0.5\textwidth]{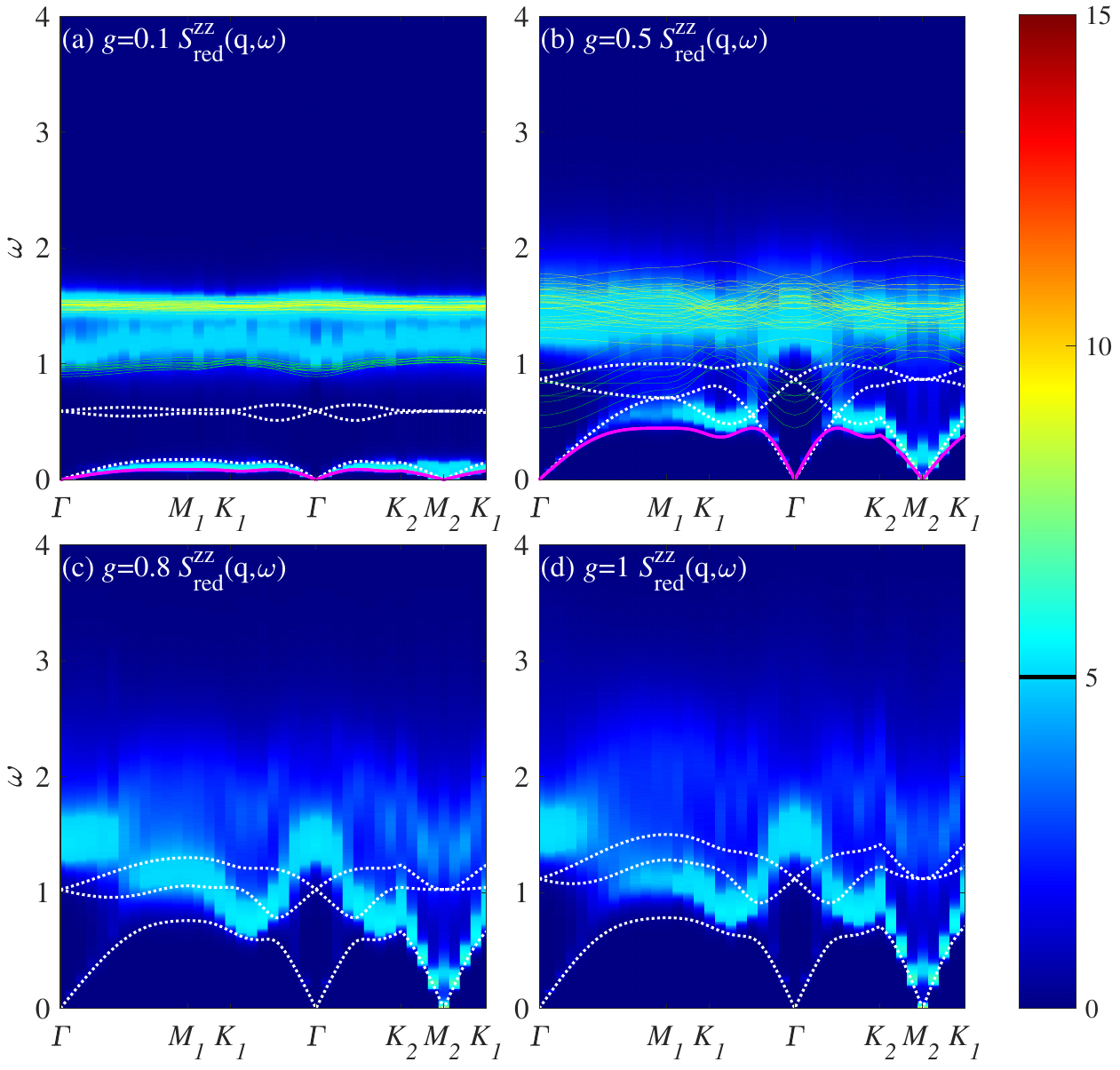}
    \caption{Dynamic spin structure factors in the reduced BZ of the trimerized Hexagon lattice at different $g$ values: (a) $g=0.1$, (b) $g=0.5$, (c) $g=0.8$, and (d) $g=1$. The pink lines illustrate the magnon dispersion obtained from LSWT using the low-energy effective block spin model. The white lines represent the LSWT results of the original model. The high-symmetry path is shown in Fig. ~\ref{lat}(d2), and we use the same logarithmic scale in SAC results when $S^{zz}_{red}(q,\omega)>U_0=5$, expressed as $U = {U_0} + {\log _{10}}{S^{zz}_{red}(q,\omega )}-{\log _{10}}{U_0}$.}
    \label{baex}
\end{figure}

\subsection{Trimerized Hexagon Lattice}
\label{hex}
Taking inspiration from the magnetic trimer structure of Ba$_4$Ir$_3$O$_{10}$ as shown in Ref. ~\cite{PhysRevLett.129.207201, cao2020quantum, cao2020quest, PhysRevB.103.224420, PhysRevB.106.075108}, the last 2D trimerized lattice we want to study is a trimerized Hexagon lattice, illustrated in Fig.~\ref{lat}(d1). Unlike the honeycomb lattice, its unit cell contains three sublattices instead of two. This lattice is also topologically equivalent to Fig.~\ref{lat}(e1). Due to the same bond dilutions as the Collinear II lattice, we observe nearly the same antiferromagnetic order in thermodynamic limit for the two lattice models, as shown in Fig.~\ref{ms2}(b). And the low-energy effective model for trimer block spins is defined on a rhombus lattice, characterized by bond exchange coupling $J_v=J_h \approx \frac{4}{9}g$, as can be seen in Fig.~\ref{eff}(d).

Fig.~\ref{baex} presents the dynamic spin structure factors in the reduced BZ. We illustrated the high-symmetry path in Fig.~\ref{lat}(d2). In the QMC-SAC spectrum at $g=0.1$, the doublon and quarton are distinct. With the increase in $g$, the magnon portion expands, the doublon and quarton quickly mix, and gradually, all three parts merge, presenting a low-energy prominent spin wave and high-energy continuum inherited from the 1D case. The yellow and green lines in Figs.~\ref{baex}(a)-(b) are the dispersions of quarton and doublon, respectively, obtained from the PA. At $g=0.5$, green lines fail to describe the intermediate energy excitation as the doublon and quarton are already mixed into the continuum spectrum. We also display the LSWT results of the original model (dotted white line) and the effective model (pink solid line). These two results accurately describe low energy magnon in two different limits of $g$, like their performance in the previous three lattices.

\begin{figure*}[htbp]
    \centering
    \includegraphics[width=1\textwidth]{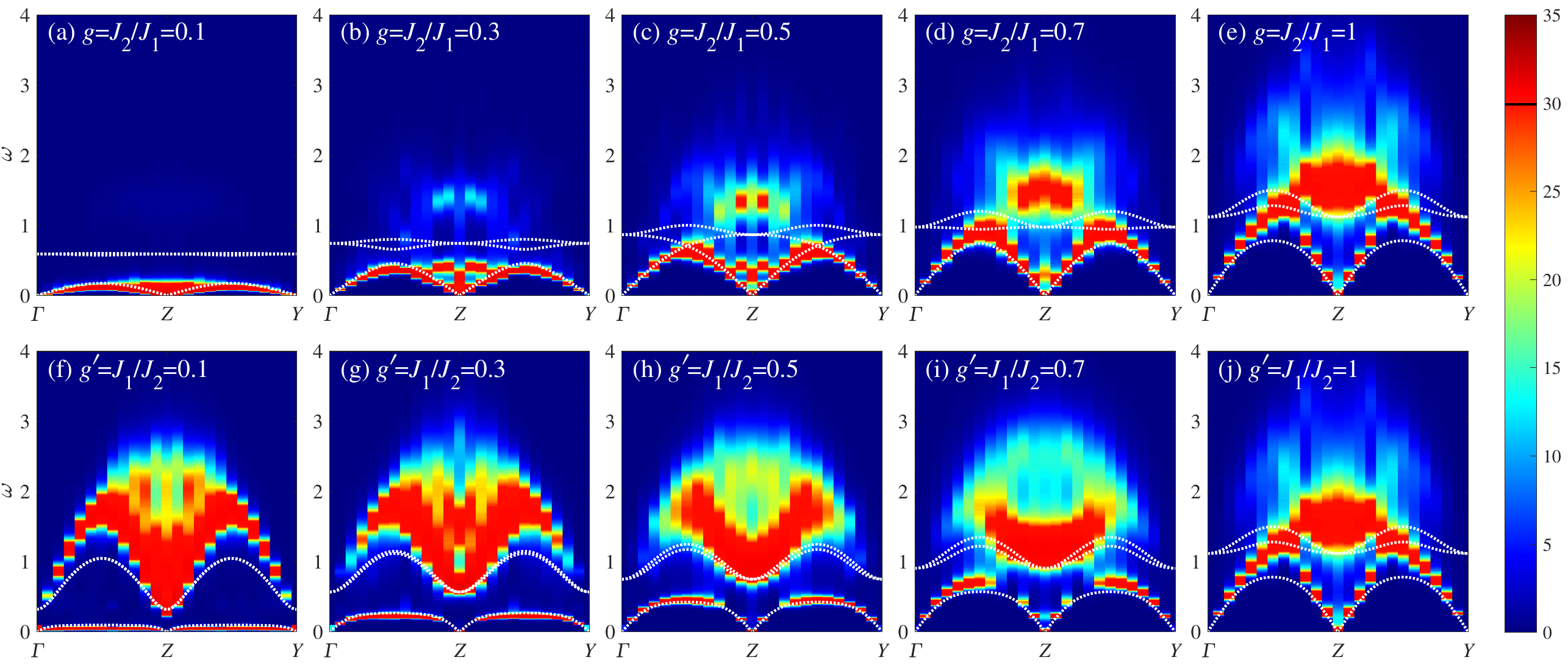}
    \caption{Dynamic spin structure factors of spin model shown in Fig. ~\ref{lat} (e1) along the path $\Gamma \rightarrow Y (0,\pi)$ at different $g$ values: (a) $g=0.1$, (b) $g=0.3$, (c) $g=0.5$, (d) $g=0.7$, (e) $g=1$, (f) $g^\prime=0.1$, (g) $g^\prime=0.3$, (h) $g^\prime=0.5$, (i) $g^\prime=0.7$, and (j) $g^\prime=1$. The white lines represent the LSWT results of the original model. The high-symmetry path is shown in Fig. ~\ref{lat}(e2), and we use the same logarithmic scale in SAC results when $S(q,\omega)>U_0=30$, expressed as $U = {U_0} + {\log _{10}}{S(q,\omega )}-{\log _{10}}{U_0}$. }
    \label{gamma}
\end{figure*}
To give a direct comparison with the experimental results of Ba$_4$Ir$_3$O$_{10}$, we used a topological equivalent lattice (see Fig. ~\ref{lat}(e1)) to study the dynamic spin structure factor along the $\Gamma(0,0)-Y(0,\pi)$ direction in the BZ of Fig. ~\ref{lat}(e2). As shown in Fig.~\ref{gamma}, the upper figures show the $S(q,\omega)$ results with varying $g=J_2/J_1$ ($J_1$ is set to 1), while the lower ones are the results with varying $g^\prime=J_1/J_2$ ($J_2=1$ is set to 1). At $g=0.1$, the spectra weight around $Z(0, \pi/2)$ is significantly strong due to the magnetic ordered ground state. With $g$ increase, the dynamic spin structure factors along the path $\Gamma \rightarrow Z \rightarrow Y$ show continuum spectra at higher energy. In addition, we can see a weak spectra weight zone between the continuum and the magnon mode. Turn to the Fig.~\ref{gamma} (f-g), when $g^\prime=J_1/J_2=0.1$, there are two branch excitations. The lower one is gapless due to the antiferromagnetic ground state, which can be effectively described by the antiferromagnetic Heisenberg model between $b$ sublattices. For the higher part, it is inherited from the two-spinon continuum of the quais-1D chain along $J_2$ bonds. As $g^\prime$ increases, the bandwidth of low-energy magnon increases, and eventually merges with the upper branch. For the RIXS spectra shown in Ref. ~\cite{PhysRevLett.129.207201}, the gapless magnon is not found in the spectrum. Its spectrum contains only the gapped or confined two-spinon continuum. In the Ba$_4$Ir$_3$O$_{10}$ material, it may have stronger intertrimer interaction $J_2$ compared to intratrimer interaction $J_1$. Some experiments have revealed the dominant 1D Luttinger liquid physics along the $J_2$ chain ~\cite{PhysRevB.92.035109, PhysRevB.89.165103}. In our case, our simulations with varying $g^\prime$ can capture the quasi-1D physics along $\Gamma \rightarrow Z \rightarrow Y$ direction. To explain the absence of low-energy magnon in the RIXS spectrum, it would be interesting to use a more sophisticated model, including easy-axis anisotropy, Dzyaloshinsky-Moriya interaction, and the possible bond randomness effect~\cite{fang2022dynamical, PhysRevB.94.174442, PhysRevB.99.085141}, to simulate the RIXS spectrum directly. We left it for future study.

\section{Analysis} \label{Analysis}
 In the previous section, we have shown some LSWT results of the effective block spin models and some perturbative dispersions for the doublon and quarton. This section will explain how we get the effective block spin models and how the perturbative analysis works for doublon and quarton excitations.

\subsection{Low Energy Effective Models and Magnons}
\label{magnon}
\begin{figure}
    \centering
    \includegraphics[width=0.48\textwidth]{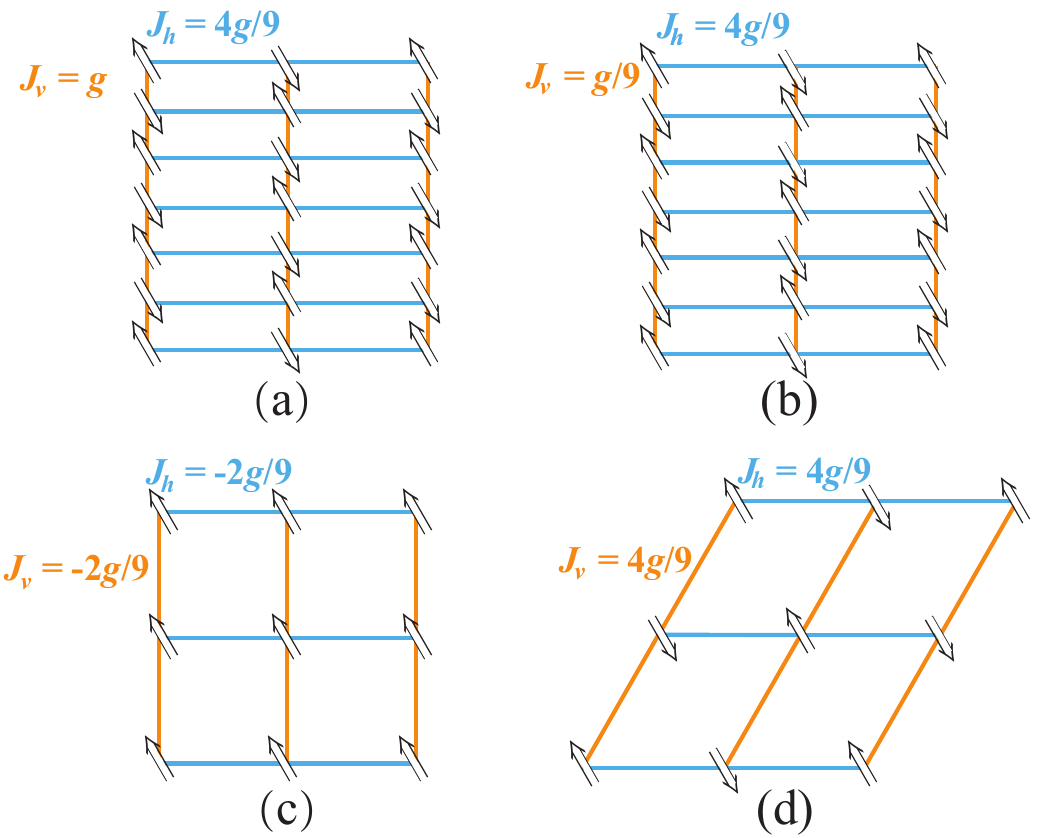}
    \caption{The low-energy effective block spin models of each 2D trimerized lattice when the value $g$ is small. (a) Collinear I lattice, (b) Collinear II lattice, (c) Trimerized Lieb lattice, (d) Trimerized Hexagon lattice. The light blue and orange lines represent the effective interactions between trimer blocks along two primitive vectors of unit cells. Here, $J_h$ represents an interaction along the horizontal direction of the primitive cell, while $J_v$ is associated with the vertical direction. For the trimerized Lieb lattice, the effective interaction is ferromagnetic. $\Uparrow$ represents the effective block spin $S=1/2$ of each trimer. The arrays of $\Uparrow$ and $\Downarrow$ show the magnetically ordered ground state of each effective model.}
    \label{eff}
\end{figure}
Due to weak coupling and the effective $S=1/2$ of each trimer, we can derive the low-energy effective block spin models of four trimerized systems, providing more insights into understanding low-energy magnon. Here, we outline the procedures for getting these effective models. Historically, the Kadanoff method has been well developed ~\cite{PhysRevB.18.3568, PhysRevLett.76.1146, PhysRevA.77.032346}, and used in the studies of quantum criticality both in the low-dimensional and high-dimensional systems ~\cite{cheng2017multipartite, PhysRevA.92.032327, PhysRevE.97.062134}. Due to the weak intertrimer interactions and an odd number of spins within a unit cell, an effective spin model can be obtained by projecting the original Hamiltonian onto the effective Hilbert space through the Kadanoff method.

We can formally split the system into the intratrimer ($H_{\rm{t}}$) and intertrimer ($H_{\rm{tt}}$) parts, which respectively contain the $J_1$ and $J_2$ couplings. To obtain the effective model, we use each trimer's two degenerate ground states to construct the basis of the low-energy effective Hilbert space. As shown in Fig.~\ref{late}, the two degenerate states are given by 
\begin{eqnarray}
&&\left|0\right\rangle^{1/2,-1/2}  = \frac{1}{{\sqrt 6 }}\left( {\left| { \uparrow  \downarrow  \downarrow } \right\rangle  - 2\left| { \downarrow    \uparrow \downarrow  } \right\rangle  + \left| { \downarrow  \downarrow   \uparrow } \right\rangle } \right), \\
&&\left|0\right\rangle^{1/2,1/2}   = \frac{1}{{\sqrt 6 }}\left( {\left| {\downarrow  \uparrow  \uparrow   } \right\rangle  - 2\left| { \uparrow  \downarrow  \uparrow } \right\rangle  + \left| {  \uparrow  \uparrow \downarrow } \right\rangle } \right),
\end{eqnarray}
For simplicity, we have changed the quantum numbers to superscripts. The first superscript is a trimer's total spin quantum number, and the second is the magnetic quantum number. These two ground states have the opposite magnetic quantum number $\pm 1/2$. We can rename the base kets in the effective Hilbert space, $\Uparrow$ and $\Downarrow$, and construct the projection operator of the $(i, j)_{th}$ trimer,
\begin{equation}
    P_{ij} = \left|0\right\rangle^{1/2,1/2}_{ij} \left\langle \Uparrow \right|_{ij}+\left|0\right\rangle^{1/2,-1/2}_{ij} \left\langle \Downarrow \right|_{ij}.
\end{equation}
where $(i, j)$ labels the 2D position of a trimer. Then, the effective Hamiltonian up to the first-order correction is given by,
\begin{equation}
\label{effectiveH}
    H_{\rm{eff}} =P^{\dagger}H_{\rm{t}} P + P^{\dagger}H_{\rm{tt}} P, 
\end{equation}
where $P= \prod_{i,j} P_{ij}$ is the total projection operator. In detail, the effective Pauli operators are obtained firstly,
\begin{equation}
    \tilde{\sigma}_{ij}^\alpha = \gamma_{ij} P_{ij}^\dagger \sigma_{ij}^\alpha P_{ij}, 
\end{equation}
where $\gamma_{ij}$ is the coefficient and $\alpha=x,y,z$. Then, inserting these effective Pauli operators into the Eq.(\ref{effectiveH}), we can obtain the effective Hamiltonian Eq.(\ref{effH}) with the effective exchange interactions shown in Fig.~\ref{eff}. The effective exchange interaction along the horizontal direction in our models is represented as $J_h$, while the effective exchange interaction along the vertical direction is denoted by $J_v$. It can be found that, for the trimerized Lieb lattice, the effective interactions along two directions are the same and have negative values, which means they are ferromagnetic interactions. While the two collinear lattices have inhomogeneous effective antiferromagnetic interactions.

Consequently, the low-energy effective Hamiltonian for these models can be reformulated to capture these nuanced characteristics
\begin{equation}
\label{effH}
  {H_{\rm{eff}}} = {J_h}\sum\limits_{i = 1}^{N} {{{\tilde S}_i} \cdot {{\tilde S}_{i + 1}}}  + {J_v}\sum\limits_{j = 1}^{{N}} {{{\tilde S}_j} \cdot {{\tilde S}_{j + 1}}}
\end{equation}

As shown in Fig.~\ref{eff}, based on these effective models, we can do the LSWT to analyze the low-energy magnon. The LSWT results are overlaid on the corresponding excitation spectra, marked with a pink line for clear visualization. Kadanoff method is exact in the $g\rightarrow 0$ limit. We have already seen that the pink solid lines match the low-energy magnons quite well in weak $g$ (for example, $g=0.1$) as shown in Figs.~\ref{coex}-~\ref{baex}. But for the Collinear I lattice, the deviation of spin wave velocity comes from the fact that there are more connecting bonds between the trimers, and a weaker $g$ is needed to see a better matching result. 
\begin{figure}
    \centering
    \includegraphics[width=0.48\textwidth]{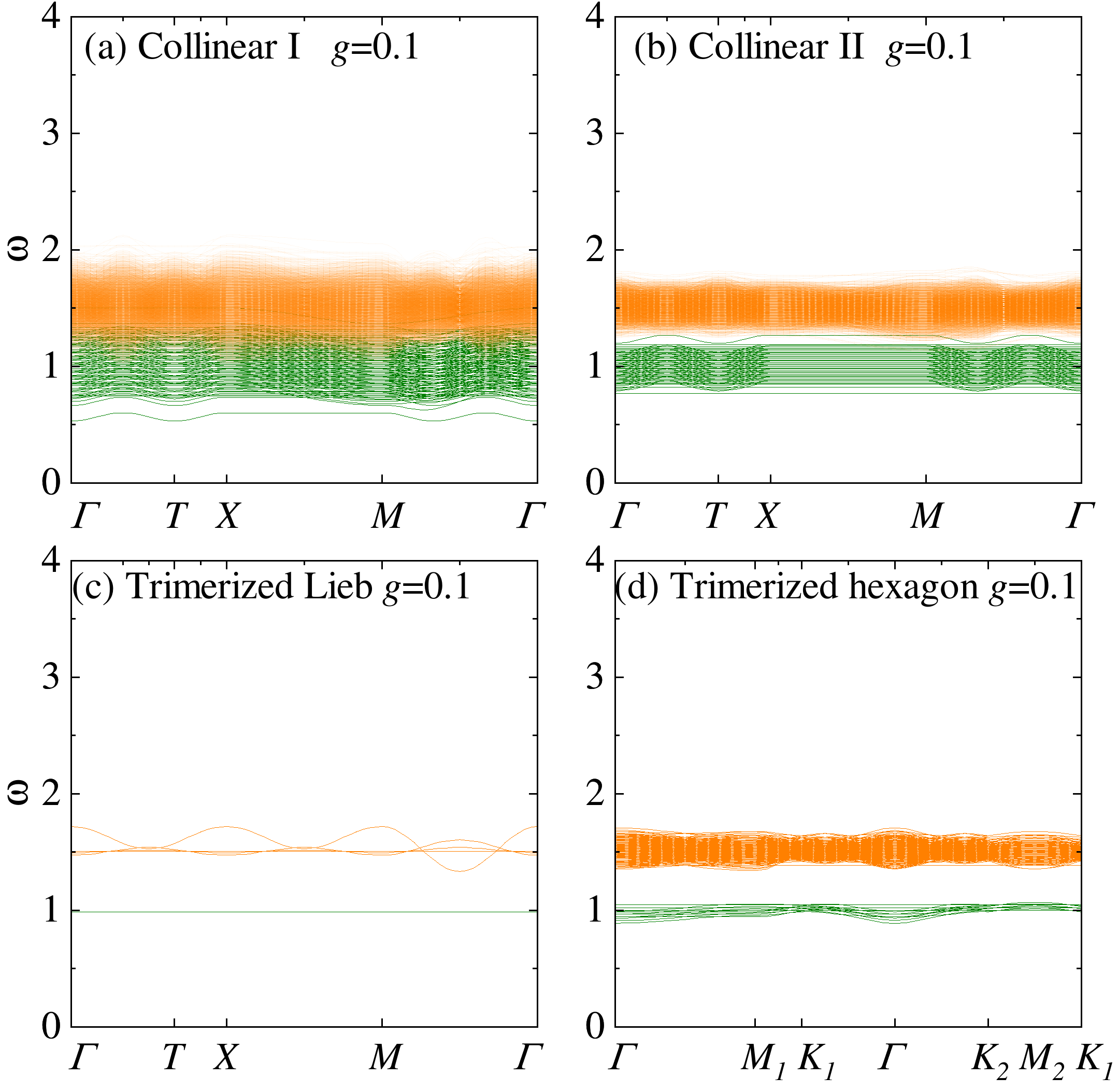}
    \caption{The perturbation dispersion relations for $g$=0.1. (a) Collinear I lattice, (b) Collinear II lattice, (c) Trimerized Lieb lattice, (d) Trimerized Hexagon lattice. The green lines denote the doublons, and the orange ones denote the quartons.}
    \label{per}
\end{figure}

\subsection{Perturbative Analysis of Doublons and Quartons}
\label{doqu}
\begin{table*}[htbp]
    \centering
    \begin{tabular}{l c c c c r}
    \hline
    Lattice  &               & excitation&   &      & dispersions\\
    \hline
 \multirow{5}{*}{Collinear I}   &      &\multirow{2}{*}{doublon} &         &  &$\epsilon =  - {E_0} + {E_1} + 7g/9$  \\
 \multirow{5}{*}{}            &           &\multirow{2}{*}{} &         &  &$\epsilon =  - {E_0} + {E_1} + 5g/9$ \\
 \cline{2-6}
 \multirow{5}{*}{}            &           &\multirow{3}{*}{quarton} &       &  &$\epsilon =  - {E_0} + {E_2} + 4g/9 + \sqrt 2 g\cos ({3q_x})/9 + \sqrt 2 g\cos ({q_y})/3$  \\ 
 \multirow{5}{*}{}            &           &\multirow{3}{*}{} &         &  &$\epsilon =  - {E_0} + {E_2} + 5g/9 + g\cos (3{q_x})/18 + \sqrt 2 g\cos ({q_y})/3$ \\
 \multirow{5}{*}{}        &               &\multirow{3}{*}{}  &        &  &$\epsilon =  - {E_0} + {E_2} + g/9 + g\cos (3{q_x})/18 + \sqrt 2 g\cos ({q_y})/9$ \\
 \hline
 \multirow{3}{*}{Collinear II} &       &doublon       &            &  &$\epsilon=  - {E_0} + {E_1} + g/9 + g\cos ({q_x})/3\sqrt 3$\\  
 \cline{2-6}
 \multirow{3}{*}{}            &            &\multirow{2}{*}{quarton}   &        &  &$\epsilon =  - {E_0} + {E_2} + g/9 + g\cos (3{q_x})/9 + 4g\cos ({q_y})/9$\\  
 \multirow{3}{*}{}            &            &\multirow{2}{*}{}  &        &  &$\epsilon =  - {E_0} + {E_2} + g/9 - g\cos (3{q_x})/6\sqrt 3  - 2g\cos ({q_y})/3\sqrt 3$\\
 \hline
 \multirow{2}{*}{Trimerized Lieb lattice} &       &doublon    &      &  &$\epsilon=  - {E_0} + {E_1} - 2g/9$\\  
 \cline{2-6}
 \multirow{2}{*}{}       &                 &quarton     &      &  &$\epsilon=  - {E_0} + {E_2} + 7g/18 - g\cos ({2q_x})/3- g\cos ({2q_y})/3$\\  
 \hline
    \end{tabular}
    \caption{The optimal dispersions in the different lattices.}
    \label{tab}
\end{table*}
The doublons and quartons in the 2D trimerized models can be conceptualized as propagating internal trimer excitations. In INS experiments and the dynamic spin structure factors, which probe the $S=1$ excitation, we focus on propagating the trimer excitation states where $|\Delta m|=1$ for doublons and $|\Delta S|=1$ for quartons. As depicted in Fig.~\ref{late}, a transition from the ground state $\left| 0 \right\rangle ^{\frac{1}{2},\frac{1}{2}}$ to the first-excited state $\left| 1 \right\rangle ^{\frac{1}{2},-\frac{1}{2}}$ results in a $|\Delta m|=1$ change which is shown by the orange dashed lines, leading to the formation of a doublon. A similar doublon excitation occurs when $\left| 0 \right\rangle ^{\frac{1}{2},-\frac{1}{2}}$ jumps to $\left| 1 \right\rangle ^{\frac{1}{2},\frac{1}{2}}$, as shown by the blue dashed lines. Quartons are more complex, representing to the second excitation states with a $|\Delta S|=1$ change, and are shown by solid lines in Fig.~\ref{late}. That is the excitation from $\left| 0 \right\rangle ^{\frac{1}{2},m}$ to $\left| 2 \right\rangle ^{\frac{3}{2},m^{\prime}}$, giving rise to the quartons. 

To further analyze the higher energy doublon and quarton excitations in weak $g$, we do the PA to give some analytical dispersions of these two quasiparticles. By ignoring the entanglement and fluctuation between trimers and regarding them as a perturbation, the ground-state wave functions of models can be seen as the product states,
\begin{equation}
 \left| \psi_g \right\rangle = \bigotimes_{i,j} \left|0 \right\rangle_{i,j}, 
\end{equation}
where $\left|0 \right\rangle_{i,j}$ is the ground state of $(i, j)_{th}$-trimer. For the Collinear I,  Collinear II, and trimerized Hexagon lattices, their low-energy effective interactions between trimer blocks are antiferromagnetic, indicating that the true ground states are the total-spin singlets satisfying $m=\sum m_{i,j}=0$. It means that the numbers of  $\left|0\right\rangle^{1/2,-1/2} $ and $\left|0\right\rangle^{1/2,1/2} $ should be equal in the ground states. For the trimerized Lieb lattice, we only choose $\left|0\right\rangle^{1/2,1/2} $ to construct the many-body ground state due to the effective ferromagnetic interaction.

In the excited state $\left| \psi \right\rangle_e$, we consider one of the $\left|0\right\rangle^{1/2,-1/2} $ and $\left|0\right\rangle^{1/2,1/2} $ is excited following the selection rule as shown in Fig.~\ref{late}.  We can calculate the dispersion relations in the reduced BZ,
\begin{eqnarray}
     \epsilon(q) &=& \left\langle H \right\rangle_e -  \left\langle H \right\rangle_g \\ \nonumber
     &=&  \left\langle \psi_e^q \right| H \left| \psi_e^q \right\rangle - \left\langle \psi_g^q \right| H \left| \psi_g^q \right\rangle, 
\end{eqnarray}
where $\left| \psi_g^q \right\rangle $ and $\left| \psi_e^q \right\rangle $ are the Fourier transformation of ground state $\left| \psi_g \right\rangle$  and excited state $\left| \psi_e \right\rangle $, respectively. In practice, we consider random arrangements of the states $\left|0\right\rangle^{1/2,-1/2} $ and $\left|0\right\rangle^{1/2,1/2} $ on each trimer to better mimic the true ground state and extract the optimal dispersion relation. We can conclude that the dispersion relations mainly depend on the excited trimers and their neighbors, only $4 \times 4$ trimers are enough to complete the derivation of dispersion relations. This method provides us with more insight to understand the intermediate-energy and high-energy excitations for the four trimerized models, which have been successfully applied in the 1D trimer chain ~\cite{cheng2022fractional}.

When the values of $g$ are small, as shown in Fig.~\ref{per}, the doublon dispersions are localized near $\omega=J_1$, while the quarton dispersions are localized near $\omega=1.5J_1$. We can also see the bandwidths of the excitation spectra. For the Collinear I lattice at $g=0.1$ in Fig.~\ref{per}(a), the dispersions seem to have already mixed. That can explain the melting of two parts of the spectrum obtained from QMC-SAC shown in Fig.\ref{coex}(a1). However, it's hard to show the spectra weight for our PA. We leave more discussion in Appendix ~\ref{SAC-test}. With $g$ increases, the doublon and quarton mix first and then merge with the low-energy magnon into magnon modes or continuum. 

We show the optimal dispersions by dashed lines in Figs.~\ref{coex}-~\ref{liebex} and show the formula in Table.~\ref{tab}, with yellow lines for the quartons and green lines for the doublons. The coefficients of $cosq_x$ and $cosq_y$ in main branches are positive in Collinear I, Collinear II, and trimerized Hexagon lattices, which correspond to the antiferromagnetic ground states. In contrast, the coefficients are negative for the effective model of trimerized Lieb lattice with a ferromagnetic ground state. We find these optimal dispersions out of thousands of dispersions. Notably, not all of the dispersions carry significant spectrum weight, and some have only a slight weight. As a result, we selectively focus on a few of the optimal dispersions, which can be used to match the CPT and QMC-SAC results closely. 

Based on the results from QMC-SAC, CPT, and PA, at small values of $g$, the magnon, doublon, and quarton are in clear distinctions. As $g$ increases, the doublon and quarton begin to mix. However, each method has its limitations. For QMC-SAC, it is hard to see the clear separation of two nearby bands with a narrow band gap. For CPT, the overestimation of magnetic order makes it difficult to describe the high-energy continuum. For PA, it only works for weak $g$. Therefore, it is challenging to determine the exact $g$ at which these three excitations start to mix. However, We can compare these results to gain a deeper understanding of the magnon, doublon, and quarton.

\section{Summary and Discussion} \label{Summary and Discussion}
\begin{figure}[t]
	\centering
	\includegraphics[width=85mm]{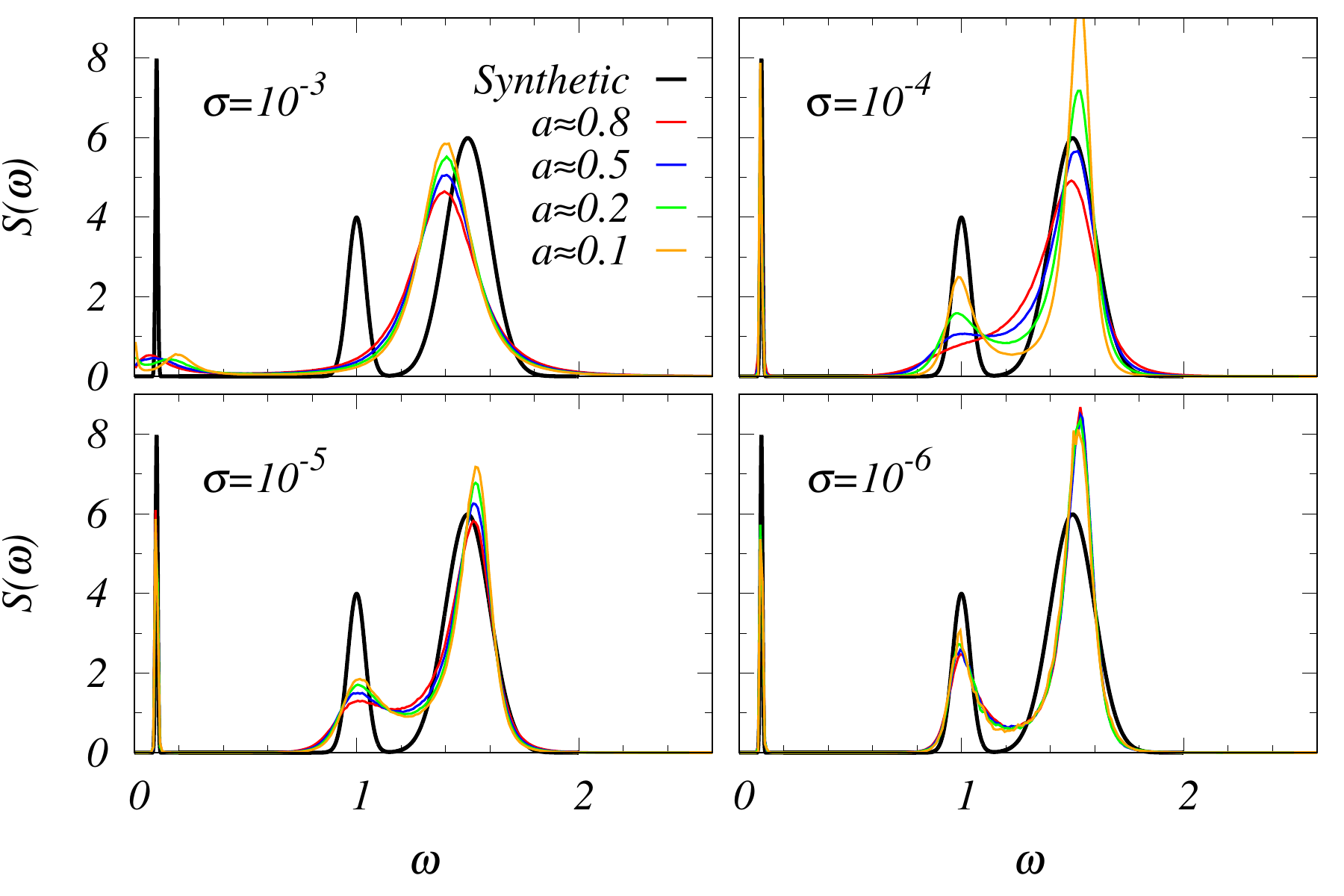}
	\caption{Test of a synthetic spectral function (black curves) using correlation data with error levels from $10^{-3}$ to $10^{-6}$. The sampling temperature is decided according to Eq. (\ref{theta-a}) with different values of $a$.}
	\label{synth}
\end{figure}

\begin{figure}[b]
	\centering
	\includegraphics[width=80mm]{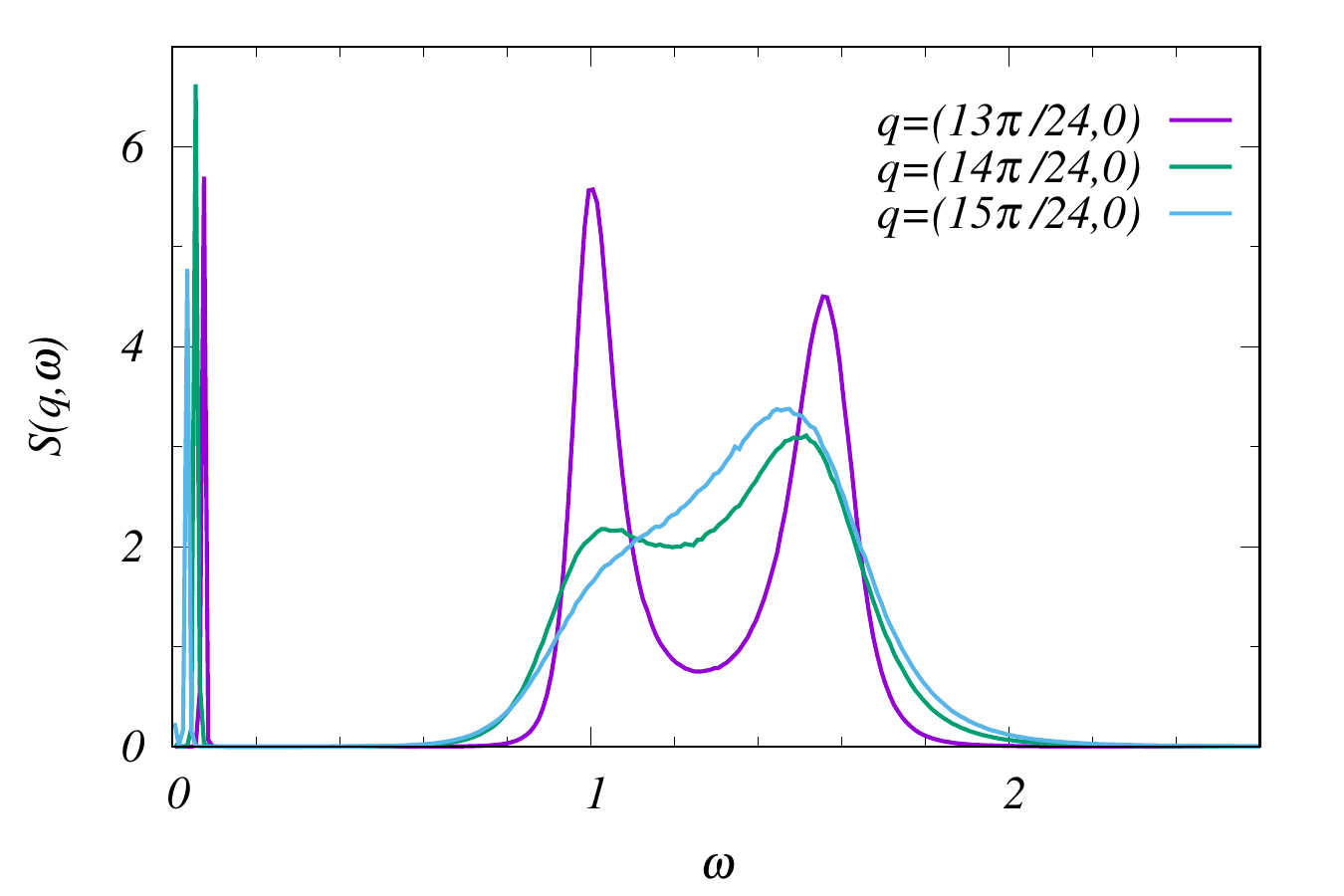}
	\caption{Spectral functions for the Collinear II model with momenta along the $ \Gamma-T$ path and close to the $T$ point.}
	\label{co2sw}
\end{figure}
In our study, we explore the evolutions of three types of excitations---magnon, doublon, and quarton for four trimerized $S=1/2$ Heisenberg models, using a combination of QMC-SAC and CPT methods with $g\in (0,1]$. In weak $g=J_2/J_1$, we can see the distinct energy separations of three quasiparticles. As the $g$ increases, the doublon and quarton bands are first to mix to form some continua. Finally, the low-energy magnon and higher-energy parts merge. For the low-energy magnon, we use linear spin wave theory for both the effective block spin model and the original model to do the analysis. The LSWT of the effective model can match the magnon curves in the weak $g$, while the LSWT of the original model can match the low-energy magnon part quite well in the large $g$ due to the strong magnetic orders of ground states. Additionally, PA is employed to qualitatively capture the behavior of doublons and quartons, particularly at small $g$ values. We provide a thorough explanation of magnons, doublons, and quartons, which enhances our understanding of how different kinds of excitations behave in 2D trimerized systems and related materials. This research expands one-dimensional trimer chains to two-dimensional trimerized systems, offering rigorous theoretical insights on 2D doublon and quarton excitations. Additionally, our numerical results and analysis explain some universal dynamic behaviors of 2D trimer systems which can help explain the patterns that will be observed in inelastic neutron scattering (INS) spectra.

In materials, a relatively weak coupling $J_3$ often exists between the $a$ and $c$ sublattices ~\cite{PhysRevLett.129.207201}. Due to the sign problem induced by the antiferromagnetic $J_3$, we can not use SSE QMC techniques to obtain the dynamic spin structure factor. Some other QMC simulations can eliminate or at least reduce the sign problem ~\cite{10.21468/SciPostPhys.12.2.054, 10.21468/SciPostPhys.3.1.005, Honecker_2022}, but it is beyond the scope of this paper. However, we believe this scheme remains unchanged even with a weak interaction $J_3$ in the $a$ and $c$ sublattices. In addition, some trimer materials, like CaNi$_3$(P$_2$O$_7$)$_2$, would have a spin quantum number larger than $1/2$ ~\cite{PhysRevB.93.184409, PhysRevB.97.224413, PhysRevB.74.024430}. For the large spin cases, such as $S=1$ and $S=3/2$, the low-energy spectrum is still magnon excitation for the 2D case, while the higher energy part is more complex, which is an interesting topic we leave for future study. 

\begin{acknowledgments}
\begin{figure}[b]
	\centering
	\includegraphics[width=80mm]{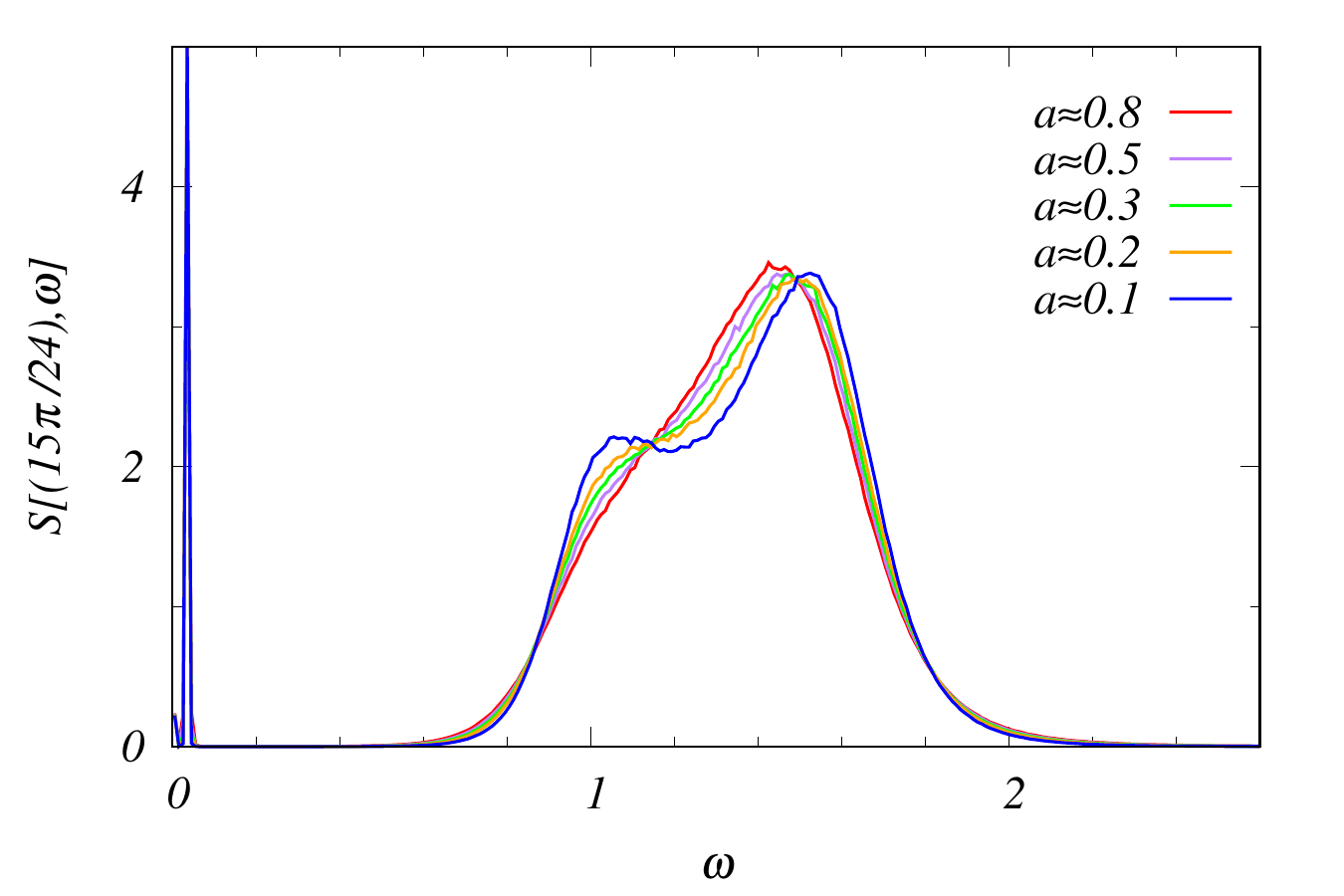}
	\caption{Spectral functions for the Collinear II model with $\boldsymbol{q}=(15\pi/24,0)$ and different sampling temperatures.}
	\label{co2q16sw}
\end{figure}

\begin{figure*}[t]
    \centering
    \includegraphics[width=1\textwidth]{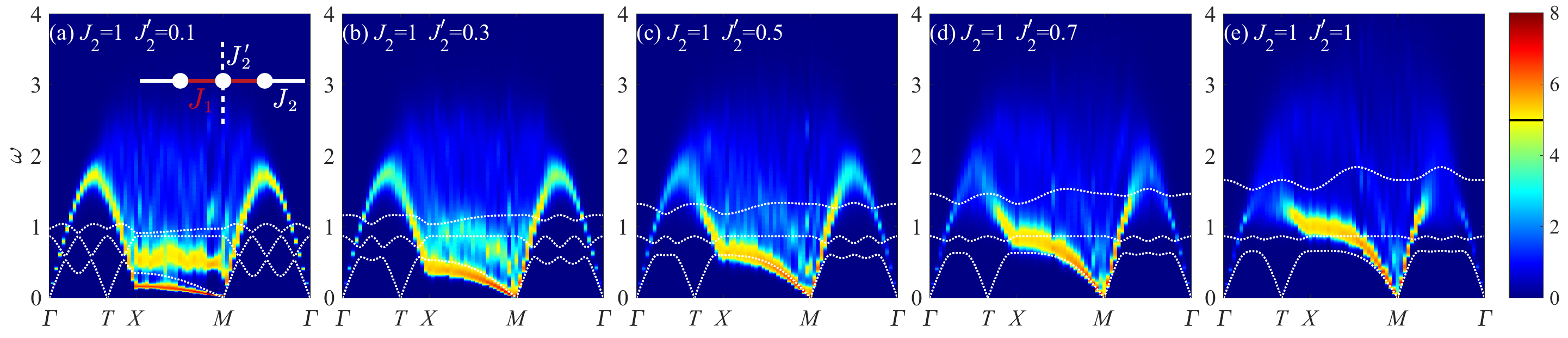}
    \caption{Dynamic spin structure factors of the trimerized Collinear II lattice at different vertical interchain $J_2^{\prime}$ interactions: (a) $J_2=1$ and $J_2^{\prime}=0.1$, we show the interactions in the inset. (b) $J_2=1$ and $J_2^{\prime}=0.3$, (c) $J_2=1$ and $J_2^{\prime}=0.5$, (d) $J_2=1$ and $J_2^{\prime}=0.7$ and (e) $J_2=1$ and $J_2^{\prime}=1$. The white lines represent the LSWT results. The high-symmetry path is shown in Fig. ~\ref{lat}(a2), and we use the same logarithmic scale in the SAC results when $S(q,\omega)>5$, expressed as $U = {U_0} + {\log _{10}}{S(q,\omega)}-{\log _{10}}{U_0}$.}
    \label{co2ic}
\end{figure*}
We would like to thank Anders W. Sandvik and Muwei Wu for the fruitful discussions. This project is supported by NKRDPC-2022YFA1402802, NSFC-11804401, NSFC-11974432, NSFC-92165204, NSFC-12047562, NSFC-12122502, Leading Talent Program of Guangdong Special Projects (201626003), and GuangDong Basic and Applied Basic Research Foundation (2023B1515120013). The calculations reported were performed on resources provided by the Guangdong Provincial Key Laboratory of Magnetoelectric Physics and Devices, No.2022B1212010008. H. Q. W also acknowledges the support from the Youth Science and Technology Talent Cultivation Project of Guangdong Provincial Association for Science and Technology. J.Q.C. also acknowledges the financial support from the Special Project in Key Areas for Universities in Guangdong Province (No. 2023ZDZX3054) and the Dongguan Key Laboratory of Artificial Intelligence Design for Advanced Materials.
\end{acknowledgments}

\appendix\vspace{0.6cm}
\section{Resolution of two close peaks in SAC} \label{SAC-test}

As presented in Fig. \ref{coex} (a1) for the Collinear I model and Fig. \ref{j3ex} (a1) for the Collinear II model, when $g=0.1$, the QMC +SAC results of the spectrum along the $ \Gamma-T$ and $M- \Gamma$ paths show either one broad high-energy peak or strong fluctuation between one and two higher-energy peaks. Despite the possible mix of the doublon and quarton excitations discussed in Sec. \ref{doqu}, another possibility is the limited performance of the SAC calculation in distinguishing two close peaks due to the relatively large error level of the QMC data.

To illustrate this issue, we first test with a synthetic spectral function which is constructed with one low-energy peak and two higher-energy peaks close to each other, as shown in Fig. \ref{synth} with the black curves. After integrating the spectral function, we add noise to the correlation data with relative error levels from $10^{-3}$ to $10^{-6}$ and apply the SAC procedure respectively. Same as in the main text, we use the unrestricted sampling method introduced in \cite{shao2023progress} with both amplitudes and frequencies updated. The fictitious sampling temperature is adjusted so that
\begin{equation}\label{theta-a}
	\left\langle\chi^2(\Theta)\right\rangle \approx \chi_{\min }^2+a \sqrt{2 \chi_{\min }^2},
\end{equation}
where $a$ is a tunable parameter. While a small value of $a$ leads to over-fitting of the correlation data and a large value of $a$  produces a featureless curve, a reasonable choice should be of order $1$ according to the $\chi^2$ distribution, typically 0.5 based on the tests in \cite{shao2023progress}. 

In Fig. \ref{synth}, we show results with $a$ from $0.1$ to $0.8$. It is clear that when the correlation data is bad, in this case with the error level $10^{-3}$, the two high-energy peaks may never be distinguished. On the other hand, when the correlation data is good enough, in this case with the error level $10^{-6}$, despite the quantitative disagreement of the peak width, the outcoming spectral function can always capture the correct spectral structure and changes very little with different $a$ values, which indicates the reliability of the result. More importantly, when the correlation data is barely good enough, in this case with the error level $10^{-4}$, the outcoming spectral function changes dramatically with different $a$ values, and the peak at $\omega=1$ can only be detected when $a\approx 0.2$ or smaller. However, if without prior knowledge of the true spectral function, this could also indicate a ``non-physical" feature due to over-fitting.

Going back to the trimerized models, In Fig. \ref{co2sw} we show a few representative spectral functions for the Collinear II model with momenta along the $ \Gamma-T$ path and close to the $T$ point, which is also shown in the color plot Fig. \ref{j3ex} (a1).  In these cases, the error level of the QMC correlation data is about $5\times 10^{-5}$, and $a=0.5$ is used to decide the sampling temperature. The obvious difference among adjacent momenta already indicates large statistical fluctuations. 

For the spectral function at $\boldsymbol{q}=(15\pi/24,0)$, similar to the test with the synthetic data, lowering the sampling temperature leads to the separation of the broad high-energy peak, as shown in Fig. \ref{co2q16sw}, and the locations of the two new peaks are around $\omega=1$ and $1.5$, which are just the doublon and quarton energy scales respectively. Even though one can not draw a concrete conclusion of what the true spectral functions look like from SAC with the current quality of our QMC data, most likely the doublon and quarton modes are still well spectated since the intertrimer interactions are very weak and the CPT calculation should be accurate in this case. 

\section{Dimensional crossover effect on Collinear II lattice}\label{dimensional}

In this section, we provide an additional explanation of the origin of the continuum spectrum at $g=1$ in the Collinear II lattice, as can be seen in Fig. ~\ref{co2ic}. In Sec.~\ref{co2} of the main text, we attribute the continuum spectra to the quasi-1D physics due to the dilute connected bonds along the $y$ direction. To clarify, we study a variant model related to Collinear II with the same $g=1$ limit. We let the interaction in the vertical or $y$ direction as $J_2^\prime$. We set $J_2=J_1=1$ and vary $J_2^\prime$. In this way, when $J_2^\prime=0$, the lattice is the decoupled AFM chains, and with $J_2^\prime$ increase, the lattice transit from 1D to 2D. We show the dynamic spin structure factors in Fig. ~\ref{co2ic}, the white lines are the LSWT results. In the quasi-1D region, shown in Figs. ~\ref{co2ic}(a)-(b), the spectra are more like two-spinon continua at path $\Gamma \rightarrow X$. With $J_2^\prime$ increase (from quasi-1D to 2D), the continuum spectra along the path $X \rightarrow M$ shift to higher energy as the single magnon mode lifts, similar to coupled spin chains model without dilution ~\cite{PhysRevLett.126.227201, yu2018deconfinement}. At $J_2^\prime$=1, We can observe that the high-energy continuum is inherited from the 1D case. A similar dimensional crossover effect can also be seen in the trimerized Hexagonal lattice.

\bibliography{ref}
\end{document}